\documentclass{amsart} 
\usepackage{graphicx}
\usepackage{amssymb, amsmath}
\usepackage{amsfonts}
\usepackage{amssymb}
\usepackage{epsfig}
\usepackage{float}

\newtheorem{theorem}{Theorem}

\newtheorem{definition}[theorem]{Definition}

\newtheorem{lemma}[theorem]{Lemma}

\newtheorem{remark}[theorem]{Remark}

\newcommand{\divv}{\text{\rm div}}

\newcommand{\cL}{\mathcal L}
\newcommand{\C}{\mathbb C}

\newcommand{\R}{\mathbb R}

\newcommand{\dist}{\text{\rm dist}}
\def\be{\begin{equation}}
\def\ee{\end{equation}}
\def\la{\lambda}

\def\R{\mathbb R}

\def\tilde{\widetilde}

\numberwithin{equation}{section}
\numberwithin{theorem}{section}
\numberwithin{figure}{section}

\begin{document}
\bibliographystyle{siam}

\title[Double-diffusive Convection]
{Bifurcation and Stability of Two-Dimensional Double-Diffusive Convection}

\author[C. Hsia]{Chun-Hsiung Hsia}
\address[CH]{Department of Mathematics, Indiana University,
Bloomington, IN 47405}
\email{chsia@indiana.edu}

\author[T. Ma]{Tian Ma}
\address[TM]{Department of Mathematics, Sichuan University,
Chengdu, P. R. China}

\author[S. Wang]{Shouhong Wang}
\address[SW]{Department of Mathematics,
Indiana University, Bloomington, IN 47405}
\email{showang@indiana.edu}

\date{\today}

\thanks{The work was supported in part by the
Office of Naval Research,  by the National Science Foundation, 
and by the National Science Foundation
of China.}


\begin{abstract}
In this article, we present a bifurcation and stability analysis on 
the double-diffusive convection. 
The main objective  is to study  1) the mechanism of 
the saddle-node bifurcation  and hysteresis for the problem, 
2) the formation, stability and transitions of the typical convection 
structures,  and 3) the stability of  solutions.
It is proved in particular  that there
are two different types of transitions: continuous and jump, 
which are determined explicitly using some physical relevant 
nondimensional parameters. It is also proved that the jump 
transition always leads to the existence of a saddle-node 
bifurcation and hysteresis phenomena.
\end{abstract}

\maketitle
\section{Introduction}
Convective motions occur in a fluid when there are density 
variations present. Double-diffusive convection is the name 
given to such convective motions when the density variations 
are caused by two different components which have different 
rates of diffusion. 
Double-diffusion was first originally discovered 
in the 1857 by Jevons \cite{jevons}, forgotten, and then rediscovered
as an ``oceanographic curiosity'' a century later; see  among others 
Stommel, Arons and 
Blanchard \cite{stommel}, Veronis \cite{gv}, 
and Baines and Gill \cite{pg}. In addition to its 
effects on oceanic circulation, double-diffusion 
convection has wide applications to  such diverse fields 
as growing crystals, the dynamics of magma chambers 
and convection in the sun.

The best known double-diffusive instabilities are ``salt-fingers'' 
as discussed in the pioneering work by Stern \cite{stern}. 
These arise when hot salty water lies over cold fresh 
water of a higher density and consist of long fingers of rising and 
sinking water. A blob of hot salty water which finds itself surrounded 
by cold fresh water rapidly loses its heat while retaining its salt due 
to the very different rates of diffusion of heat and salt. The blob 
becomes cold and salty and hence denser than the surrounding fluid. 
This tends to make the blob sink further, drawing down more hot 
salty water from above giving rise to sinking fingers of fluid. 

The main objective of this article is to develop a 
bifurcation and stability theory for the double-diffusive convection, 
including

\begin{enumerate}

\item[1)] existence of bifurcations/transitions, 

\item[2)] asymptotic stability of bifurcated solutions, 
and 

\item[3)] the structure/patterns and their stability/transitions 
in the physical space.

\end{enumerate} 

The analysis is based on a  bifurcation theory 
for nonlinear partial differential equations 
and a geometric  theory of two-dimensional (2D) incompressible flows,  
both developed recently by two of the authors; see  respectively 
\cite{b-book, mw-db1} and \cite{amsbook} and the references therein.

This bifurcation theory is centered at a new notion of bifurcation, called 
attractor bifurcation for dynamical systems, both finite dimensional and 
infinite dimensional. The main ingredients of the theory include 
a) the attractor bifurcation theory, b) steady state bifurcation for a 
class of nonlinear problems with even order non-degenerate 
nonlinearities, regardless of the multiplicity of the 
eigenvalues, and c) new strategies for 
the Lyapunov-Schmidt reduction and the center manifold 
reduction procedures. The bifurcation theory has been applied to 
various problems from science and engineering, including, in particular, 
the Kuramoto-Sivashinshy equation, the Cahn-Hillard equation, 
the Ginzburg-Landau equation, Reaction-Diffusion equations 
in Biology and Chemistry, the B\'enard convection and the Taylor problem in 
fluid dynamics.

The geometric theory of 2D incompressible flows 
was initiated by the authors 
to study the structure and its stability and transitions
of 2-D incompressible fluid flows in the physical spaces. 
This program of study  
consists of research in directions: 
1) the study of the structure and its transitions/evolutions 
of divergence-free vector fields, and 2) the study of 
the structure and its transitions 
of velocity fields for 2-D incompressible fluid flows governed 
by the Navier-Stokes equations or the Euler equations.
The study in Area 1)  is more kinematic in nature, and the results and methods 
developed can naturally be applied to other problems of mathematical physics 
involving divergence-free vector fields.
In fluid dynamics context, the study in  Area 2) involves specific 
connections between the solutions of the Navier-Stokes or the Euler 
equations and flow structure in the physical space. In other words, 
this area of research links the kinematics to the dynamics of 
fluid flows. This is unquestionably an important and difficult problem.
Progresses have been made in several directions. First, 
a new rigorous characterization of 
boundary layer separations for 2D viscous incompressible flows is developed 
recently by the authors, in collaboration in part 
with Michael Ghil; see \cite{amsbook} and the references therein. 
Another example in this area is the structure (e.g. rolls) in the 
physical space in the Rayleigh-B\'enard convection, using the structural 
stability theorem developed in Area 1) together 
with the application of the aforementioned bifurcation theory; 
see \cite{benard, amsbook}.

In this article, we consider two-dimensional 
double-diffusive convections modelled by the Boussinesq equations with 
two diffusion equations of the temperature and salinity functions.
In comparison to the  Rayleigh-B\'enard convection case, 
and the steady linearized problem around the basic state 
for the double-diffusive convection 
is nonsymmetric. This leads to a much harder eigenvalue problem, 
and consequently  much more involved bifurcation and stability analysis.
Hence, the bifurcation and the flow structure are much richer.

The central gravity of the analysis 
is the reduction of the problem 
to the center manifold in the first unstable eigendirections, 
based on an approximation formula for the center manifold function. 
The key idea is to find  the  approximation of the 
reduction to certain order, leading to a  ``nondegenerate'' system 
with higher order perturbations. The full bifurcation and stability 
analysis are then carried out 
using a combination of  
the attractor bifurcation theory and the 
geometric theory of 2D incompressible flows. 

We now address briefly the main characteristics of the 
bifurcation and stability analysis of the two-dimensional 
double-diffusive model presented in this article.  

\medskip

{\sc First}, 
the double-diffusive system involves four important nondimensional 
parameters:  the thermal Rayleigh number
$\lambda$, the solute Rayleigh number $\eta$, the Prandtl number $\sigma$
and the Lewis number $\tau$, defined by (\ref{parameter}). 
We examine in this article different transition/instability regimes defined by 
these parameters. We aim to get a better understanding of 
the different physical mechanisms involved in the onset of convection.
It is hoped that this will enable progress to be made in the 
theoretical understanding of the onset of double 
diffusive instabilities.

From the physical point of view, it is natural to consider only the case 
where the Prandtl number $\sigma>1$. 
The Lewis number $\tau$ measures the  ratio of two diffusivities. 
From the oceanic circulation point of view, the
heat diffuses about 100 times more rapidly than salt \cite{stern}; 
hence $\tau <1$. 
In this case, different regimes of stabilities and instabilities/transitions 
of the basic state can be described by regions in the $\lambda$-$\eta$ plane
(the thermal and salt Rayleigh numbers) as shown in 
Figure~\ref{fig4.1}. In this article, we focus on the regimes where 
\be \label{eq1.1}
\eta <  \eta_c= \frac{27}{4}\pi^4\tau^2(1+\sigma^{-1})(1-\tau)^{-1}.\ee
In the case where $\eta >  \eta_c$, transitions to periodic or aperiodic 
solutions are expected, and will be addressed elsewhere.

\medskip

{\sc Second}, 
we show that there are two different 
transition regimes: continuous and jump, dictated  by a 
nondimensional parameter 
\be \label{eq1.2}\eta_{c_1}=\frac{27}{4}\pi^4\tau^3(1-\tau^2)^{-1}.\ee

For the regime with $\eta < \eta_{c_1}$, the transition is continuous 
when the thermal Rayleigh number $\lambda$ crosses  a critical value
\be \label{eq1.3} \lambda_c(\eta)= \frac{\eta}{\tau}+\frac{27}{4}\pi^4.
\ee
The rigorous result in this case is stated in Theorem~\ref{th2.3}.

For the regime with $\eta_{c_1}< \eta < \eta_c$, 
the transition is jump near $\lambda=\lambda_c(\eta)$. 
It is shown also that for this case, there is a saddle-node bifurcation 
at $\lambda^\ast(\eta) <\lambda_c(\eta)$, together with 
the hysteresis feature of the transition in 
$\lambda^\ast(\eta) <\lambda < \lambda_c(\eta)$. 
The rigorous result in this case is stated in Theorem~\ref{th2.4}, and 
schematically illustrated  by Figure~\ref{fg2.2}.

\medskip

{\sc Third}, as an attractor, the bifurcated attractor has 
asymptotic stability  in the sense that it attracts 
all solutions with initial data in the phase space outside of the 
stable manifold of the basic state.
As Kirchg\"assner indicated in  \cite{kirch}, 
an ideal stability theorem would include all physically meaningful 
perturbations and establish the local stability of a selected class of 
stationary solutions, and today we are still far from this goal.
On the other hand, fluid flows are normally time dependent. Therefore  
bifurcation analysis for steady state problems provides in general 
only partial answers to the problem, and is not enough for solving 
the stability problem. Hence it appears that the right notion of 
asymptotic stability should be best described 
by the attractor near, but excluding, the trivial state. It is one of 
our main motivations for introducing the attractor bifurcation.

\medskip

{\sc Fourth}, another important aspect of the study is 
to classify the structure/pattern of the
solutions after the bifurcation. A natural tool to attack this problem is the 
structural stability of the solutions in the physical space. Thanks 
to the aforementioned geometric theory of 2D incompressible flows, 
the structure and its transitions of the convection states
 in the physical space 
is analyzed, leading in particular to a rigorous justification of the 
roll structures. More patterns and structures associated with the double 
diffusive models will be studied elsewhere. 

\medskip

{\sc Fifth}, for mathematical completeness and for applications
 to other physical problems, the bifurcation analysis is also carried out
for the case where the Lewis number $\tau>1$. In this case, only the 
continuous transition is present. The rigorous result in this case 
is stated in Theorem~\ref{th2.5}.  

\medskip

This article is organized as follows.
The basic governing equations are given in Section 2, and 
the main theorems are stated in Section 3. The remaining sections 
are devoted to the proof of the main theorems, with Section 4 
on a recapitulation of the attractor bifurcation theory and the geometric 
theory of incompressible flows, Section 5 on eigenvalue problems, Section 
6 on center manifold reductions, and Section 7 on the completion of the 
proofs.

\section{Equations and Set-up}
\subsection{Boussinesq equations}
In this paper, we consider the double-diffusive convection 
problem in  a two-dimensional (2D) domain  $\mathbb R^1 \times 
(0, h) \subset \R^2$ ($h > 0$) with coordinates denoted by $(x, z)$.
 The Boussinesq equations, which govern the motion and states 
of the fluid flow, are as follows; see  Veronis 
\cite{gv}: 
\be
\label{eq2.1}
\left\{
\begin{aligned}
& \frac{\partial U}{\partial t}+(U \cdot \nabla)U
=-\frac{1}{\rho_0}(\nabla p + \rho g e)+\nu \Delta U,\\
& \frac{\partial T}{\partial t}+(U \cdot \nabla)T=
\kappa_T \Delta T,\\
& \frac{\partial S}{\partial t}+(U \cdot \nabla)S=
\kappa_S \Delta S,\\
&\divv \,\, U = 0,
\end{aligned}
\right.
\ee 
where $U=(u,w)$  is the velocity function,
$T$  is the temperature function, 
$S$ is  the solute concentration, 
$P$ is  the pressure,
$g$ is  the gravity constant, 
$e=(0, 1)$  is the unit vector in the $z$-direction, 
the constant  $\nu>0$  is the kinematic viscosity, 
the constant  $\kappa_T>0$  is the thermal diffusity, 
the constant  $\kappa_S>0$  is the solute diffusity,
the constant  $\rho_0>0$  is the fluid density at the lower surface $z=0$,
and $\rho$ is the fluid density given by the following equation of state
\begin{equation}
\label{eq2.2}
\rho=\rho_0[1-a(T-T_0)+b(S-S_0)]. 
\end{equation}
Here  $a$  and  $b$ are assumed to be positive constants. 
Moreover, the lower
boundary ($z=0$) is maintained at a constant temperature $T_{0}$ and
a constant solute concentration $S_{0}$, while the upper boundary ($z=h$)
 is maintained at a constant temperature $T_{1}$ and 
a constant  solute concentration
 $S_{1}$,  where  $T_{0}>T_{1}$ and $S_{0}>S_{1}$. The trivial steady state 
 solution of (\ref{eq2.1}) - (\ref{eq2.2}) is given by
\be
\label{eq2.3}
\left\{
\begin{aligned}
& U^{0}=0,\\
& T^{0}=T_{0}-(\frac{T_{0}-T_{1}}{h})z,\\
& S^{0}=S_{0}-(\frac{S_{0}-S_{1}}{h})z,\\
& p^{0}=p_{0}-g\rho_{0}[z+\frac{a}{2}(\frac{T_{0}-T_{1}}{h})z^{2}-\frac{b}{2}
(\frac{S_{0}-S_{1}}{h})z^{2}], 
\end{aligned}
\right.
\ee
where $p_{0}$ is a constant.
To make the equations non-dimensional, we consider the perturbation of 
the solution from the 
trivial convection state:
\begin{align*}
& U''= U-U^{0}, && \qquad  T''= T-T^{0},\\
& S''= S-S^{0},&& \qquad  p''= p-p^{0}.
\end{align*}
Then we set
\begin{align*}
& (x, z)= h(x', z'),  && \qquad t= h^{2}t'/\kappa_{T}, \\
& U'' = \kappa_{T}U'/h, && \qquad T''=(T_{0}-T_{1})T', \\
& S'' = (S_{0}-S_{1})S',  && \qquad p''
 = \rho_{0}\nu\kappa_{T}p'/h^{2}.
\end{align*}
Omitting the primes, the equations (\ref{eq2.1}) can be 
written as 
\be
\label{eq2.4}
\left\{
\begin{aligned}
& \frac{\partial U}{\partial t}=\sigma( \Delta U-\nabla p)
  +\sigma(\lambda T-\eta S)e-(U \cdot \nabla)U,\\
& \frac{\partial T}{\partial t}=\Delta T +w-(U \cdot \nabla)T,\\
& \frac{\partial S}{\partial t}=\tau \Delta S+w- (U \cdot \nabla)S,\\
& \divv  U = 0,
\end{aligned}
\right.
\ee
for $(x,z)$ in the non-dimensional domain $\Omega=\mathbb R^{1}\times 
(0,1)$, where $U=(u,w)$,  and the positive nondimensional 
parameters used above are given by 
\be
\label{parameter}
\left\{
\begin{aligned}
& \lambda=\frac{ag(T_{0}-T_{1})h^{3}}{\kappa_{T}\nu} && \quad \text{the
 thermal Rayleigh number},\\
& \eta=\frac{bg(S_{0}-S_{1})h^{3}}{\kappa_{T}\nu}  &&\quad \text{the salinity
Rayleigh number},\\
& \sigma= \frac{\nu}{\kappa_{T}}  &&\quad \text{the Prandtl number},\\ 
& \tau=\frac{\kappa_{S}}{\kappa_{T}} &&\quad \text{the Lewis number}. 
\end{aligned}\right.
\ee
We consider periodic boundary condition in the $x$-direction 
\be
\label{eq2.5}
(U, T, S)(x,z,t)=(U, T, S)(x+2k\pi/\alpha,z,t) \quad \forall k\in \mathbb Z.
\ee
At the top and bottom boundaries, we impose 
 the free-free boundary conditions given by 
\be
\label{eq2.6}
(T, S, w)=0, \qquad \frac{\partial u}{\partial z}=0 
\quad \text{at} \quad z=0,1. 
\ee 
It's natural to put the constraint
\begin{equation}
\label{eq2.7}
\int_{\Omega} udxdz=0,
\end{equation}
for the problem (\ref{eq2.4})-(\ref{eq2.6}).
It is easy to see from the following 
computation that  (\ref{eq2.4}) is invariant under this constraint:
\begin{align*}
 \frac{\partial}{\partial t}\int_{0}^{1}\int_{0}^{2\pi/\alpha}u dx dz =&
\int_{0}^{1}\int_{0}^{2\pi/\alpha}\sigma (u_{xx}+u_{zz}-p_{x})
      -(uu_{x}+wu_{z})dxdz\\
       =&\int_0^1\sigma (u_x-p)\mid_{x=0}^{x=2\pi/\alpha}dz 
+ \int_{0}^{2\pi/\alpha}\sigma (u_z\mid_{z=0}^{z=1})dx\\
        &+\int_0^1\int_0^{2\pi/\alpha}(u_x+w_z)udxdz\\
        =&0.
\end{align*} 

The initial value conditions are given by
\be
\label{eq2.10}
(U,T,S)=(\tilde{U},\tilde{T},\tilde{S}) \quad  \text{at}\quad t=0.
\ee

\subsection{Functional setting}

Let 
\begin{align*}
 H=& \{(U,T,S)\in L^{2}(\Omega)^{4}\quad \mid \quad 
      \divv \, U=0, w\mid_{z=0,1}=0, 
      \int_{\Omega} u dxdz=0,\\
  & u \text{ is $2\pi/\alpha$-periodic in $x$-direction } \},\\
 V=& \{(U,T,S)\in H^{1}(\Omega)^{4}\cap H \,\quad |\quad \,(U,T,S)\,\,
    \text{is $2\pi/\alpha$-periodic}
    \text{ in x-direction, }\\
   & \,\, T\mid_{z=0,1}=S\mid_{z=0,1}=0  \},\\
 H_{1}=& V\cap H^{2}(\Omega).
\end{align*}
Let $G: H_{1} \to H$ 
and $L_{\lambda \eta}=-A-B_{\lambda \eta}: H_{1} \to H$ 
be defined by
\begin{align*}
& G (\psi) = (- P[(U\cdot\nabla )U], 
                       - (U\cdot \nabla )T, 
                       - (U\cdot \nabla )S ),  \\
& A\psi = ( -P [\sigma(\Delta U)], -\Delta T, -\tau \Delta S ), \\
& B_{\lambda \eta} \psi =  (-P [\sigma(\lambda T-\eta S)e],-w,-w ), 
\end{align*} 
for any $\psi=(U, T, S) \in H_1$. Here $P$ is the Leray projection to 
$L^2$ fields. 
Then the Boussinesq equations (\ref{eq2.4})-(\ref{eq2.7}) can be written 
in the following operator form
\begin{equation}\label{eq2.11}
\frac{d\psi}{dt} = L_{\lambda \eta} \psi + G(\psi), \qquad \psi=(U,T,S).
\end{equation}

\section{Main Results}  

\subsection{Definition of attractor bifurcation}
In order to state the main theorems of this article, 
we proceed with the definition of attractor bifurcation, first introduced by 
two of the authors in \cite{mw-db1,b-book}.

Let  $H$ and  $H_1$ be two Hilbert spaces,
and $H_1 \hookrightarrow H$ be a dense and compact inclusion.
We consider the following
nonlinear evolution equations
\be
\label{eq3.1}
\left\{
\begin{aligned}
& \frac{du}{dt} = L_\lambda u +G(u,\lambda), \\
& u(0) = u_0,
\end{aligned}
\right.
\ee
where $u: [0, \infty) \to H$  is the unknown function, $\lambda \in
\mathbb R$  is the  system  parameter, and
$L_\lambda:H_1\to H$ are parameterized linear completely
continuous fields depending continuously on $\lambda\in \R^1$, which
satisfy
\begin{equation}
\label{eq3.2}
\left\{\begin{aligned}
& -L_\lambda = A + B_\lambda && \text{a sectorial operator}, \\
& A:H_1 \to H && \text{a linear homeomorphism}, \\
& B_\lambda :H_1\to H && \text{parameterized linear compact
operators.}
\end{aligned}\right.
\end{equation}
It is easy to see \cite{henry} that $L_\lambda$
generates an analytic semi-group $\{e^{tL_\lambda}\}_{t\ge 0}$.
Then we can define  fractional power operators $(-L_\lambda)^{\mu}$ for any
$0\le \mu \le 1$ with domain $H_\mu = D((-L_\lambda)^{\mu})$ such that
$H_{\mu_1} \subset H_{\mu_2}$ if $\mu_1 > \mu_2$, and $H_0=H$.

Furthermore, we assume that the nonlinear terms
$G(\cdot, \lambda):H_\mu \to H$ for some $1> \mu \ge 0$
are a family of parameterized $C^r$
bounded operators ($r\ge 1$) continuously depending on the parameter
$\lambda\in \R^1$, such that
\begin{equation}
\label{eq3.3}
 G(u,\lambda) = o(\|u\|_{H_\mu}), \quad \forall\,\, \lambda\in \R^1.
\end{equation}

In this paper, we are interested in the sectorial operator
$-L_\lambda = A +B_\lambda$ such
 that there exist an eigenvalue sequence $\{\rho_k\}
\subset \C^1$ and an eigenvector sequence $\{e_k, h_k\}\subset
H_1$ of $A$:
\begin{equation}
\label{eq3.4}
\left\{\begin{aligned}
& Az_k = \rho_kz_k,  \qquad z_k=e_k + i h_k, \\
& \text{Re} \rho_k\to \infty \,\,(k\to\infty), \\
& |\text{Im} \rho_k / (a + \text{Re} \rho_k) | \le c,
\end{aligned}\right.
\end{equation}
for some $a, c > 0$, such that
$\{e_k, h_k\}$ is a basis of $H$.

Condition (\ref{eq3.4})  implies that $A$ is a sectorial operator.
For the operator $B_\lambda:H_1\to H$, we also assume that
there is a constant $0<\theta<1$ such that
\begin{equation}
\label{eq3.5}
B_\lambda :H_\theta \longrightarrow H \,\,\text{bounded, $\forall$
$\lambda\in \R^1$.}
\end{equation}
Under conditions (\ref{eq3.4}) and (\ref{eq3.5}), the operator
$-L_\lambda=A + B_\lambda$ is a sectorial operator.

Let $\{S_\lambda(t)\}_{t\ge 0}$ be an operator semi-group generated by
the equation (\ref{eq3.1}).
Then the solution of (\ref{eq3.1})  can be expressed
as 
\[
\psi(t, \psi_0) = S_\lambda(t)\psi_0, \qquad t\ge 0.
\]
\begin{definition}
\label{df3.4}
A set $\Sigma \subset H$ is called an invariant set of
(\ref{eq3.1}) if $S(t) \Sigma = \Sigma$ for any $t\ge 0$. An
invariant set $\Sigma \subset H$ of (\ref{eq3.1}) is said to be an
attractor if $\Sigma$ is compact, and there exists a neighborhood $W
\subset H$ of $\Sigma$ such that for any $\psi_0\in W$ we have 
$$
\lim_{t\to \infty}\dist_H(\psi(t,\psi_0),\Sigma)= 0.
$$
\end{definition}

\begin{definition}
\label{df3.5}
\begin{enumerate}

\item We say that the solution to equation (\ref{eq3.1}) bifurcates 
from $(\psi,\lambda) =
(0,\lambda_0)$ to an invariant set $\Omega_\lambda$, if there exists a
sequence of invariant sets $\{\Omega_{\lambda_n}\}$ of (\ref{eq3.1})
such that $0 \notin \Omega_{\lambda_n}$, and
\begin{align*}
& \lim_{n\to \infty} \lambda_n = \lambda_0,  \\
& \lim_{n\to \infty} \max_{x\in \Omega_{\lambda_n}} |x| =0. 
\end{align*}

\item If the invariant sets $\Omega_\lambda$ are attractors of
(\ref{eq3.1}), then the bifurcation is called attractor bifurcation.

\end{enumerate}
\end{definition}

\subsection{Main theorems}
We now consider the double diffusive equations (\ref{eq2.4}). 
In this article, we always consider the case 
where the parameters $\lambda$ 
and $\eta$ satisfying
\begin{equation}
\label{eq2.9}
\begin{aligned}
 & \eta < \eta_c=\frac{27}{4}\pi^4\tau^2(1+\sigma^{-1})(1-\tau)^{-1},\\
 & \lambda \approx \lambda_c= \frac{\eta}{\tau}+\frac{27}{4}\pi^4.
\end{aligned}     
\end{equation}

First we consider a more physically relevant diffusive regime  
where the thermal Prandtl number $\sigma$ 
is bigger than $1$, and the Lewis number $\tau$ is less than $1$:
\begin{equation}
\label{eq2.8}
 \sigma> 1 >\tau, \,\,  \alpha^2=\pi^2/2. 
\end{equation}
Here the condition on $\alpha$ defines the aspect ratio of the domain.
In this case, we consider two straight lines in the 
$\lambda-\eta$ parameter plane as shown in Figure~\ref{fig4.1}:
\begin{figure}
 \centering \includegraphics[height=.35\hsize]{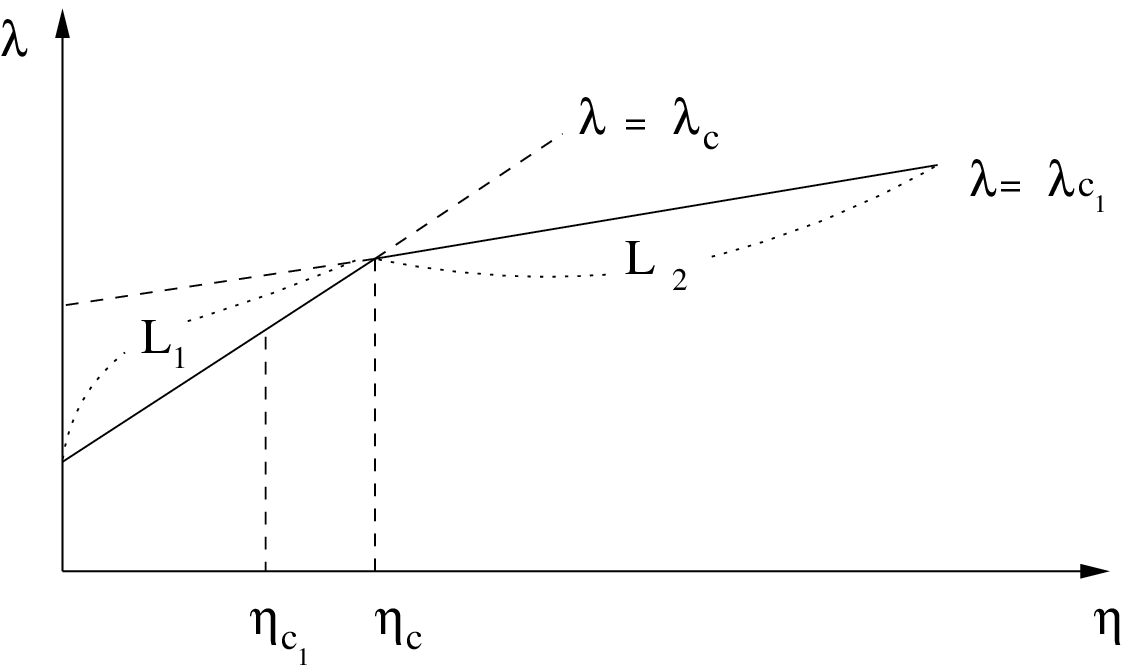}
 \caption{}
\label{fig4.1}
\end{figure}
\begin{equation}
\left\{
\begin{aligned}
& L_1: \quad \lambda=\lambda_c(\eta), \\
& L_2: \quad \lambda=\lambda_{c_1}(\eta), 
\end{aligned}
\right.
\end{equation}
where 
\begin{equation}
\left\{
\begin{aligned}
& \lambda_c(\eta)= \frac{\eta}{\tau}+\frac{27}{4}\pi^4, \\
& \lambda_{c_1}(\eta) = \frac{(\sigma+\tau)}{(\sigma+1)}\eta+
      \frac{27}{4}\pi^4(1+\sigma^{-1}\tau)(1+\tau),
\end{aligned}
\right.
\end{equation}
Also shown in Figure~\ref{fig4.1} are two critical values for $\eta$ given by 
$$\eta_c=\frac{27}{4}\pi^4\tau^2(1+\sigma^{-1})(1-\tau)^{-1}, \quad 
\eta_{c_1}=\frac{27}{4}\pi^4\tau^3(1-\tau^2)^{-1}.$$

The following two main theorems study the transitions/bifurcation 
of the double-diffusive model near the line $L_1$ for 
$\eta < \eta_c$.

\begin{theorem}
\label{th2.3}
Assume that the condition (\ref{eq2.8}) holds true, and
$\eta<\eta_{c_1}=\frac{27}{4}\pi^4\tau^3(1-\tau^2)^{-1}$. Then the following 
assertions for the problem (\ref{eq2.4})-(\ref{eq2.7})  hold true.
\begin{enumerate}
\item If $\lambda \le \lambda_c$, the steady state $(U,T,S)=0$ is
      locally asymptotically stable for the problem.
\item The solutions bifurcate from $((U,T,S),\lambda)=(0,\lambda_c)$ to 
an attractor $\Sigma_\lambda$ for $\lambda>\lambda_c$, which
 is homeomorphic to $S^{1}$, and consists of steady state solutions  
of the problem.  
\item For any $\psi_0=(\tilde{U},\tilde{T},\tilde{S})\in H\backslash\Gamma$,
      there exists a time $t_0 \ge 0$ such that for any $t \ge t_0 $, the
      vector field $U(t,\psi_0)$ is topologically equivalent to the structure 
      as shown in Figure~\ref{fg2.3} , 
      where $\psi=(U(t,\psi_0),T(t,\psi_0),S(t,\psi_0))$
      is a solution of (\ref{eq2.4})-(\ref{eq2.7}) with 
      (\ref{eq2.8})-(\ref{eq2.9}), $\Gamma$ is the stable manifold of 
      the trivial solution $(U,T,S)=0$ with co-dimension 2 in $H$.
 \end{enumerate} 
\end{theorem}

\begin{figure}
 \centering \includegraphics[height=.4\hsize]{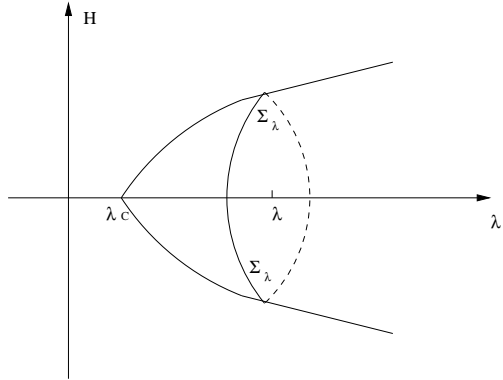}
 \caption{If $\tau>1$ or $\eta<\eta_{c_1}$, the equations
          bifurcate from $(0,\lambda_c)$ to an attractor 
         $\Sigma_{\lambda}$ for $\lambda>\lambda_c$. }
 \label{fig2.1}
\end{figure}

\begin{theorem}
\label{th2.4}
Assume that the condition (\ref{eq2.8}) holds true, and
$\eta_c>\eta>\eta_{c_1}=\frac{27}{4}\pi^4\tau^3(1-\tau^2)^{-1}$. Then  there
exists a saddle-node bifurcation point $\lambda_0$ ($\lambda_0<\lambda_c$)
for the equations, such that the following statements for the problems
(\ref{eq2.4})-(\ref{eq2.7}) hold true.
\begin{enumerate}
\item At $\lambda=\lambda_0$, there is an invariant set 
$\Sigma_0=\Sigma_{\lambda_0}$ with $0 \notin \Sigma_0 $.
\item For $\lambda<\lambda_0$, there is  no invariant set near $\Sigma_0$.
\item For $\lambda_0<\lambda<\lambda_c$, there are two branches of invariant
      sets $\Sigma_{\lambda}^{1}$ and $\Sigma_{\lambda}^{2}$, and 
      $\Sigma_{\lambda}^{2}$ extends to $\lambda\ge\lambda_c$ and near 
      $\lambda_c$ as well.
\item For each $\lambda>\lambda_0$, $\Sigma_{\lambda}^{2}$ 
      is an attractor with $ dist(\Sigma_{\lambda}^2,0)>0$.
\item For $\lambda_0<\lambda<\lambda_c$,
     \begin{enumerate}
     \item  $\Sigma_{\lambda}^{1}$ is a repeller with  
           $ 0\notin \Sigma_{\lambda}^{1}$, and
     \item  when $\lambda$ is near $\lambda_c$, $\Sigma_{\lambda}^{1}$ is
         homeomorphic to $S^1$, consisting of steady states.
     \end{enumerate} 
\end{enumerate}
\end{theorem}
\begin{figure}
 \centering \includegraphics[height=.4\hsize]{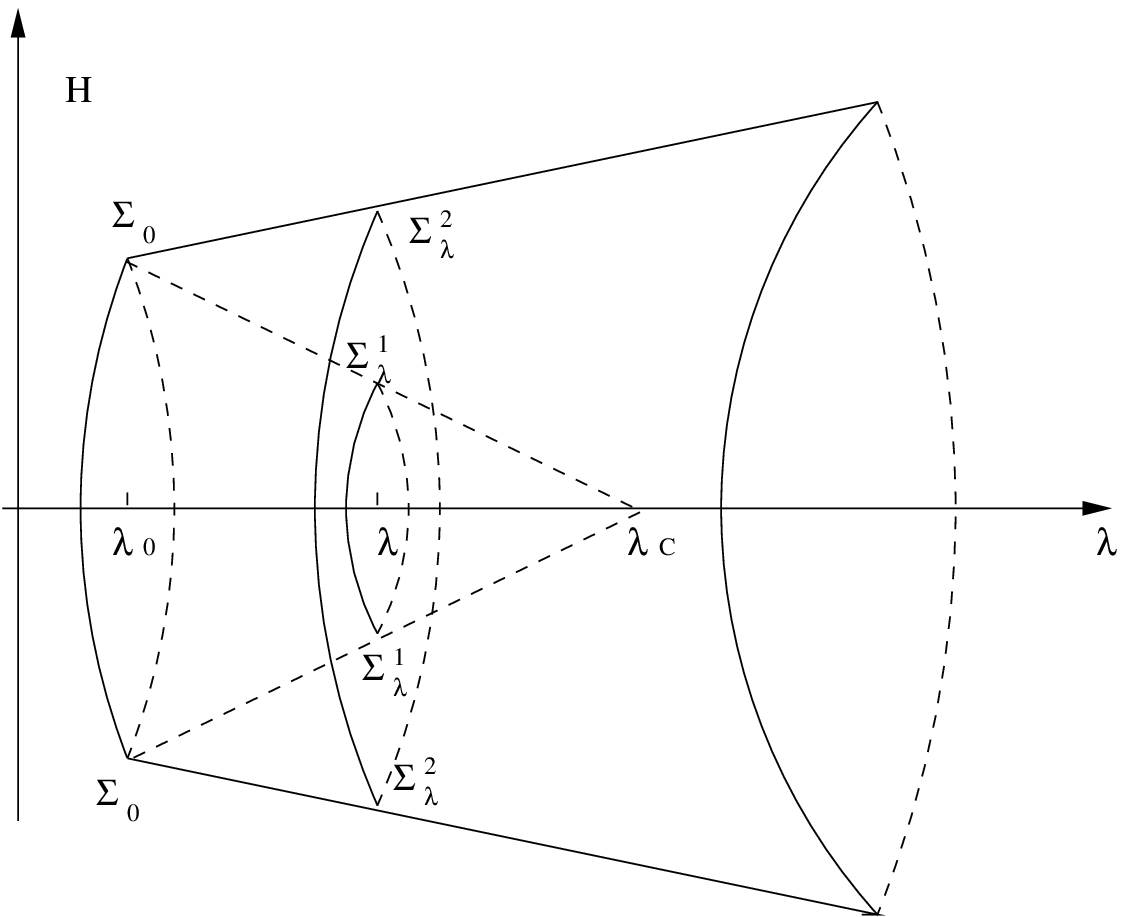}
 \caption{If $\tau<1$ and $\eta_{c_1}<\eta<\eta_c$, the equations
        have a saddle-node bifurcation for $\lambda<\lambda_c$.}
 \label{fg2.2}
\end{figure}

\begin{figure}
 \centering \includegraphics[height=.4\hsize]{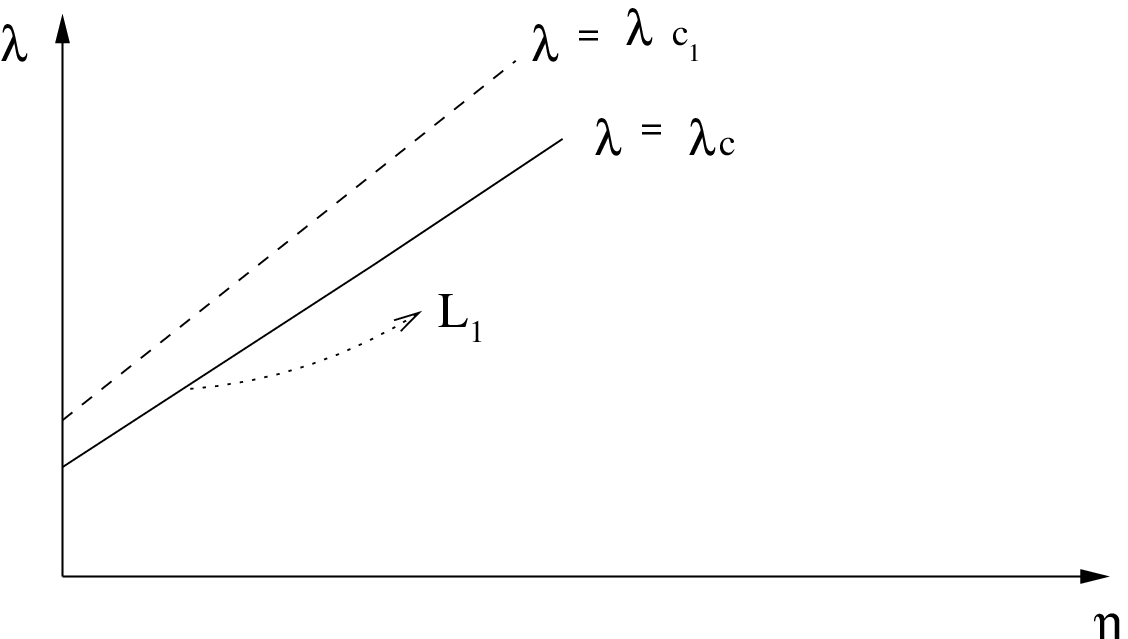}
 \caption{}
 \label{fig4.2}
\end{figure}

\begin{figure}
 \centering \includegraphics[height=.18\hsize]{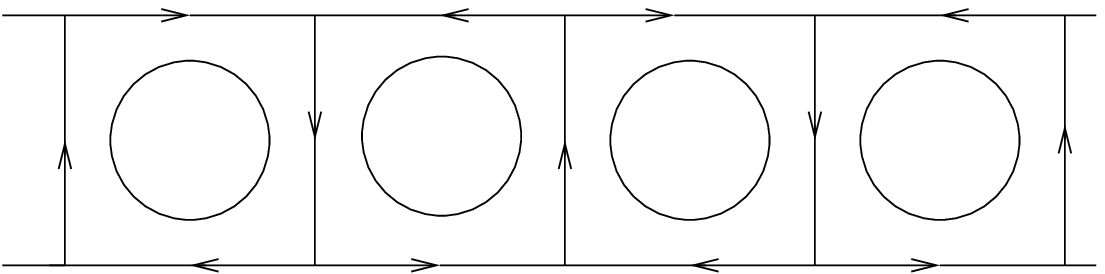}
 \caption{}
 \label{fg2.3}
\end{figure}

We now consider the diffusive parameter regime where 
$\sigma>1$, $\tau>1$, $\alpha^2=\pi^2/2$ and $\sigma\ne \tau$. In this case, 
two lines are shown in Figure~\ref{fig4.2}.
The followin theorem provides bifurcation when $\lambda$ 
crosses the line $L_1$.

\begin{theorem}
\label{th2.5}
Assume that  $\sigma>1$, $\tau>1$, 
  $\alpha^2=\pi^2/2$, $\sigma\ne \tau$  and
 (\ref{eq2.9}) hold, then for any $\eta>0$, the 
 following assertions 
for the problem  (\ref{eq2.4})-(\ref{eq2.7}) hold true.
\begin{enumerate}
\item If $\lambda < \lambda_c$, the steady state $(U,T,S)=0$ is
      locally asymptotically stable for the problem.
\item The solutions bifurcate from $((U,T,S),\lambda)=(0,\lambda_c)$ to 
an attractor $\Sigma_\lambda$ for $\lambda>\lambda_c$, which is 
homologically equivalent to $S^{1}$, and consists of steady state solutions  
of the problem.  
\item For any $\psi_0=(\tilde{U},\tilde{T},\tilde{S})\in H\backslash\Gamma$,
      there exists a time $t_0 \ge 0$ such that for any $t \ge t_0 $, the
      vector field $U(t,\psi_0)$ is topologically equivalent to the structure 
      as shown in Figure~\ref{fg2.3} , 
      where $\psi=(U(t,\psi_0),T(t,\psi_0),S(t,\psi_0))$
      is a solution of (\ref{eq2.4})-(\ref{eq2.7}), 
$\Gamma$ is the stable manifold of 
      the trivial solution $(U,T,S)=0$ with co-dimension 2 in $H$.
 \end{enumerate} 
\end{theorem}

\section{Preliminaries}
\subsection{Attractor bifurcation theory}
Consider (\ref{eq3.1}) satisfying (\ref{eq3.2})  and (\ref{eq3.3}). 
We start with the Principle of Exchange of Stabilities (PES).
Let the eigenvalues (counting the multiplicity) of $L_\lambda$ be given by
\[
\beta_1(\lambda),\beta_2(\lambda),\cdots,\beta_k(\lambda),\cdots \in \mathbb C.
\]

Suppose that
\begin{equation}
\label{eq3.13} Re\beta_i(\lambda)\begin{cases}
<0 \,\,\,\, \text{if}\,\,\,\, \lambda<\lambda_0\\
=0 \,\,\,\, \text{if}\,\,\,\, \lambda=\lambda_0\\
>0 \,\,\,\, \text{if}\,\,\,\, \lambda>\lambda_0
\end{cases}
\qquad(1\le i\le m)
\end{equation}

\begin{equation}
\label{eq3.14} Re\beta_j(\lambda_0)<0.\qquad \forall \,\,m+1\le j.
\end{equation}

Let the eigenspace of $L_\lambda$ at $\lambda_0$ be
\[
E_0=\displaystyle{\bigcup_{1\le j \le m}\bigcup_{k=1}^{\infty}}\{u,v\in H_1\mid
(L_{\lambda_0}-\beta_j(\lambda_0))^k w=0, w=u+iv \}.
\]
It is known that $\dim E_0=m$.

\begin{theorem}[T. Ma and S. Wang \cite{mw-db1,b-book}]
\label{th3.6} 
Assume that the conditions (\ref{eq3.2})-(\ref{eq3.5}) and
(\ref{eq3.13})-(\ref{eq3.14}) hold true, 
and $u=0$ is locally asymptotically 
stable  for (\ref{eq3.1}) 
at $\lambda=\lambda_0$. Then the following assertions hold true.

\begin{enumerate}
\item (\ref{eq3.1}) bifurcates from $(u,\lambda)=(0,\lambda_0)$ to 
attractors $\Sigma_\lambda$, having the same homology as $S^{m-1}$, 
for $\lambda>\lambda_0$, 
with $ m-1\le dim\Sigma_\lambda\le m$, which is connected 
as $ m>1$;

\item For any $u_\lambda\in\Sigma_\lambda$, $u_\lambda$ can 
be expressed as
\[
u_\lambda=v_\lambda+o(\|v_\lambda\|_{H_1}), \,\, v_\lambda\in E_0;
\]
\item There is an open set $U\subset H$ with $0\in U$ such that the 
attractor $\Sigma_\lambda$ bifurcated from $(0,\lambda_0)$ 
attracts $U\backslash\Gamma$ in $ H$, where $\Gamma$ is the stable 
manifold of $u=0$ with co-dimension m.
\end{enumerate}
\end{theorem}

In the case where $m=2$, the bifurcated attractor can be further classified.
Consider a two-dimensional system as follows:
\be
\label{eq3.15}
\frac{dx}{dt}=\beta(\la) x - g(x, \la), \quad x \in \mathbb R^2.
\ee
Here $\beta(\la)$  is a continuous function of $\la$ 
satisfying 
\be
\label{eq3.16}
\beta(\la)\left\{
\begin{aligned}
& < 0 && \quad \text{ if } \la < \la_0, \\
& = 0 && \quad \text{ if } \la = \la_0, \\
& > 0 && \quad \text{ if } \la > \la_0, 
\end{aligned}
\right.
\ee
and 
\be
\label{eq3.17}
\left\{
\begin{aligned}
& g(x, \la)=g_k(x, \la) + o(|x|^k), \\
& g_k(\cdot, \la) \text{ is a k-multilinear field}, \\
& C_1 |x|^{k+1} \le <g_k(x, \la), x> \le C_2 |x|^{k+1},
\end{aligned}
\right.
\ee
for some integer $k=2m+1 \ge 3$, and some constants $C_2 > C_1 > 0$.

The following theorem was proved in \cite{b-book}, which shows that 
under conditions (\ref{eq3.16})  and (\ref{eq3.17}), the system 
(\ref{eq3.15}) bifurcates to an $S^1$-attractor.

\begin{theorem}
\label{th3.7}
Let the conditions (\ref{eq3.16})  and (\ref{eq3.17}) hold true. Then 
the solution to the system  (\ref{eq3.15}) 
bifurcates from $(x, \la)=(0,\la_0)$ to an attractor
$\Sigma_\la$ for $\la > \la_0$, which is homeomorphic to $S^1$. Moreover,
one and only one of the following is true.

\begin{enumerate}
\item $\Sigma_\la$ is a periodic orbit,

\item $\Sigma_\la$ consists of only singular points, or

\item $\Sigma_\la$ contains at most $2(k+1)=4(m+1)$ singular points, and
has $4N + n$ ($N+n\ge 1$) singular points, $2N$ of which are saddle points,
$2N$ of which are stable node points (possibly degenerate), and $n$ of which
have index zero.

\end{enumerate}
\end{theorem}

\subsection{Center manifold reduction}
A crucial ingredient for the proof of the main theorems using the above 
attractor bifurcation theorems is an approximation 
formula for center manifold functions derived in \cite{b-book}.

Let $H_1$  and $H$ be decomposed into
\begin{equation}
\label{eq3.6} \left\{
\begin{aligned}
& H_1 = E^\lambda_1 \oplus E^\lambda_2,  \\
& H = \widetilde E^\lambda_1 \oplus \widetilde E^\lambda_2,
\end{aligned}\right.
\end{equation}
for $\la$ near $\lambda_0 \in \R^1$, where $E^\lambda_1$,
$E^\lambda_2$ are invariant subspaces of $L_\lambda$, such that
\begin{align*}
&\dim E^\lambda_1<\infty, \\
& \tilde E^\lambda_1 = E^\lambda_1, \\
& \widetilde E^\lambda_2 = \text{closure of $E^\lambda_2$ in $H$.}
\end{align*}
In addition,  $L_\lambda$ can be decomposed into $L_\lambda =
\cL^\lambda_1 \oplus \cL^\lambda_2$ such that for any $\lambda$
near $\lambda_0$,
\begin{equation}
\label{eq3.7}
\begin{cases}
\cL^\lambda_1 = L_\lambda |_{E^\lambda_1} : E^\lambda_1
\longrightarrow \widetilde E^\lambda_1, & \\
\cL^\lambda_2 = L_\lambda|_{E^\lambda_2}:E^\lambda_2 \longrightarrow
\widetilde E^\lambda_2, &
\end{cases}
\end{equation}
where all eigenvalues of $\cL^\lambda_2$ possess negative real
parts, and the eigenvalues of $\cL^\lambda_1$ possess  nonnegative
real parts at $\lambda=\lambda_0$.

Thus, for $\lambda$ near $\lambda_0$, equation (\ref{eq3.1}) can be
written as
\begin{equation}
\label{eq3.8} 
\left\{
\begin{aligned}
& \frac{dx}{dt} = \cL^\lambda_1 x +G_1(x,y,\lambda), & \\
& \frac{dy}{dt} = \cL^\lambda_2 y + G_2(x,y,\lambda), &
\end{aligned}
\right.
\end{equation}
where $u=x+y \in H_1$, $x\in E^\lambda_1$, $y\in E^\lambda_2$,
$G_i(x,y,\lambda) = P_iG(u,\lambda)$, and $P_i:H\to \widetilde E_i^\la$
are canonical projections.
Furthermore, let
$$E^\la_2(\mu)=\text{ closure of $E^\la_2$ in } H_\mu,
$$
with $\mu < 1$ given by (\ref{eq3.3}).

By the classical center manifold theorem (see among others 
\cite{henry,temam}), 
 there exists a neighborhood of 
$\lambda_0$ given by $|\lambda-\lambda_0|<\delta$ for some 
$\delta>0$, a neighborhood $B_\lambda \subset E^\lambda_1$ of 
$x=0$, and a $C^1$ center manifold 
function $\Phi(\cdot,\lambda):B_\lambda \to 
E^\lambda_2(\theta)$, called the center manifold function, 
depending continuously on $\lambda$.
Then to investigate the
dynamic bifurcation of (\ref{eq3.1}) it suffices to consider the
finite dimensional system as follows
\begin{equation}
\label{eq3.9}
\frac{dx}{dt} = \cL^\lambda_1 x + g_1(x,\Phi_\lambda(x),\lambda),
\qquad x\in B_\lambda \subset E^\lambda_1.
\end{equation}

Let the nonlinear operator $G$ be in the following form
\begin{equation}
\label{eq3.10}
G(u,\lambda)  = G_k(u,\lambda) + o(\|u\|^k),
\end{equation}
for some integer $k \ge 2$. Here $G_k$ is a $k$-multilinear operator
\begin{align*}
&G_k:H_1 \times \cdots \times H_1 \longrightarrow H, \\
&G_k(u,\lambda) = G_k(u,\cdots,u,\lambda).
\end{align*} 

\begin{theorem}\cite{b-book}
\label{th3.2}
Under the conditions (\ref{eq3.6}), (\ref{eq3.7}) and (\ref{eq3.10}), the
center manifold function $\Phi(x,\lambda)$ can be expressed as 
\begin{equation}
\label{eq3.11}
\Phi(x,L) = (-\cL^\lambda_2)^{-1} P_2G_k(x,\lambda) + o(\|x\|^k) +
O(|\text{\rm Re}\beta|\, \|x\|^k), 
\end{equation}
where $\cL^\lambda_2$ is as in (\ref{eq3.7}), $P_2:H\to 
\widetilde E_2$ the canonical projection, $x\in E^\lambda_1$, and 
$\beta= (\beta_1(\lambda),\cdots,\beta_m(\lambda))$ the 
eigenvectors of $\cL^\lambda_1$. 
\end{theorem}

\begin{remark}
\label{rm3.3}
{\rm
Suppose that  $\{e_j\}_{j}$, the (generalized) eigenvectors of
  $L_\lambda$, form a basis of $H$ with 
  the dual basis $\{e_j^*\}_j$  such that 
$$
(e_i,e_j^*)_H \begin{cases}
&=0   \quad \text{if} \quad i\ne j,\\
&\ne 0 \quad  \text{if} \quad i=j.
\end{cases}
$$
 Then, we have 
\begin{align*}
& u=x+y \in E_1^\la \oplus E_2^\la,\\
& x = \sum^m_{i=1} x_i e_i \in E_1^\la, \\
& y = \sum^{\infty}_{i=m+1} x_i e_i \in E_2^\la.
\end{align*}
Hence, near $\la=\la_0$, $P_2G_k(x,\lambda)$ can be expressed as follows. 
\be 
\label{eq3.12}
P_2G(x, \la) = \sum^{\infty}_{j=m+1} G_k^j(x, \la) e_j 
 + o(\|x\|^k), \ee
where 
\begin{align*}
& G_k^j(x, \la)= 
\sum_{1\le j_1, \cdots, j_k \le m} a_{j_1 \cdots j_k}^j 
x_{j_1}\cdots x_{j_k}, \\
& a_{j_1 \cdots j_k}^j = (G_k(e_{j_1}, \cdots,e_{j_k}, \la), e_j^*)_H
/(e_j,e_j^*). 
\end{align*}
In many applications, the coefficients $a_{j_1 \cdots j_k}^j$ can be computed, 
and the first $m$ eigenvalues $\beta_1(\la), \cdots, \beta_m(\la)$
satisfy 
$$|\text{\rm Re}\beta(\la_0)| =
\sqrt{\sum^m_{j=1}(\text{\rm Re}\beta_j(\la_0))^2}=0.
$$
Hence (\ref{eq3.11}) and (\ref{eq3.12})  give an explicit 
formula for the first approximation of the 
center manifold functions.
}
\end{remark}

\subsection{Structural stability theorems}
In this subsection, we recall some results on structural stability 
for 2D divergence-free vector fields developed in \cite{amsbook}, 
which are 
crucial to study the asymptotic structure in the physical space 
of the bifurcated  solutions of the double-diffusive problem.

Let $C^r(\Omega,\R^2)$ be the space of all $C^r$ $(r\ge 1)$ vector
fields on $\Omega=\mathbb R^1 \times (0, 1)$, 
which are periodic in $x$ direction with period $2\pi/\alpha$, 
let $D^r(\Omega,\R^2)$ be the space of all $C^r$  
divergence-free vector
fields on $\Omega=\mathbb R^1 \times (0, 1)$, 
which are periodic in $x$ direction with period $2\pi/\alpha$, 
and with no normal 
flow condition in $z$-direction:
\[
D^r(\Omega, \R^2) 
= \left\{ v=(u,w)(x,z)\in C^r(\Omega,\R^2)\mid w = 0 \text{ at } z=0, 1\ \right\}.
\]
Furthermore, we let 
\begin{align*}
& B^r_0(\Omega, \R^2) 
= \left\{ v\in D^r(\Omega,\R^2)\mid v = 0 \text{ at } z=0, 1\ \right\}, \\
& B^r_1(\Omega, \R^2) 
= \left\{ v\in D^r(\Omega,\R^2)\quad \Big|\quad  
\begin{matrix}
\displaystyle  v = 0 \text{ at } z=0 \\
\displaystyle  w=\frac{\partial u}{\partial z}=0 \text{ at } z=1
     \end{matrix} \ \right\}.
\end{align*}

\begin{definition}
\label{df3.8}
Two vector fields $v_1,v_2 \in C^r(\Omega, \R^2)$ are called topologically
equivalent if there exists a homeomorphism of $\varphi:\Omega\to
\Omega$, which takes the orbits of $v_1$ to orbits of $v_2$ and preserves
their orientation.
\end{definition}

\begin{definition}
\label{df3.9}
Let $X= D^r(\Omega, \R^2)$  or $X= B^r_0(\Omega, \R^2)$. 
A vector field $v_0\in X$ is called structurally stable
in $X$ if there exists a neighborhood $U\subset X$ of $v_0$ 
such that for any $v\in U$, $v_0$ and $v$ are
topologically equivalent.
\end{definition}

Let $v\in D^r(\Omega, \R^2)$. 
We recall next some basic facts and definitions on divergence--free
vector fields. 
\begin{enumerate}
\item A point $p\in \Omega$ is called a singular point of $v$ if
$v(p) =0$; a singular point $p$ of $v$ is called non-degenerate if
the Jacobian matrix $Dv(p)$ is invertible; $v$ is called regular if
all singular points of $v$ are non-degenerate.

\item An interior non-degenerate singular point of $v$ can be either
a center or a saddle, and a non-degenerate boundary singularity must
be a saddle.

\item Saddles of $v$ must be connected to saddles. An interior saddle
$p\in \Omega$ is called self-connected if $p$ is connected only to
itself, i.e., $p$ occurs in a graph whose topological form is that of
the number 8.
\end{enumerate}

\begin{theorem}
\label{th3.10}
For vector fields satisfying free-free boundary conditions, we set
\begin{align*}
& B^r_2(\Omega, \R^2) 
= \left\{ v\in D^r(\Omega,\R^2)\ \Big| \  
w=\frac{\partial u}{\partial z}=0 \text{ at } z=0, 1\ \right\}, \\
& B^r_3(\Omega, \R^2) 
= \left\{ v\in B^r_2(\Omega,\R^2)\quad \Big|\quad \int_{\Omega} u dxdz =0\ 
\right\}.
\end{align*}
Then 
$v\in B^r_2(\Omega, \R^2)$ (resp.$v\in B^r_3(\Omega, \R^2)$) 
is structurally stable in $B^r_2(\Omega, \R^2)$ (resp. 
in $B^r_2(\Omega, \R^2)$) if and only if
\begin{enumerate}
\item[1)]  $v$ is regular;

\item[2)]  all interior saddle points of $v$ are self-connected; and

\item[3)]  each boundary saddle of $v$ is connected to boundary saddles 
on the same connected component of $\partial \Omega$ (resp. 
each boundary saddle of $v$ is connected to boundary saddles
not necessarily on the same connected component).
\end{enumerate}
\end{theorem}

\section{Eigenvalue problem}
In order to apply the center manifold theory to reduce the 
bifurcation problems, we shall analyze the following eigenvalue 
problem for the linearized equations of (\ref{eq2.4})-(\ref{eq2.7}). 
\be
\label{eq4.1}
\left\{
\begin{aligned}
& \sigma(\Delta U- \nabla p)+(\sigma \lambda T-\sigma \eta S)e=\beta U,\\
& \Delta T + w =\beta T,\\
& \tau \Delta S + w = \beta S,\\
& \divv U = 0,\\
& \frac{\partial u}{\partial z}\mid_{z=0,1}=w\mid_{z=0,1}=
T\mid_{z=0,1}=S\mid_{z=0,1}=0.
\end{aligned}
\right.
\ee 
We prove couples of lemmas to show that the operators $-L_{\lambda \mu}$ 
are sectorial operators when the parameters are properly chosen. In order to 
 get the precise form of the center manifold reduction, 
the eigenspaces are analyzed in detail in this section.

\subsection{Eigenvalues}
We shall use the method of separation of variables to deal with
problem (\ref{eq4.1}). Since $\psi=(U,T,S)$ is periodic 
in $x$-direction with period 
$2\pi/\alpha$, we expand the fields in Fourier series as
\be
\label{eq4.2}
\psi(x,z)=\sum_{j=-\infty}^{\infty}\psi_{j}(z)e^{ij\alpha x}.
\ee 

Plugging (\ref{eq4.2}) into (\ref{eq4.1}), we obtain the following 
system of ordinary differential equations
\be
\label{eq4.3}
\left\{
\begin{aligned}
& D_j u_j - ij\alpha p_j=\sigma^{-1}\beta u_j,\\
& D_j w_j-p_j'+\lambda T_j - \eta S_j= \sigma^{-1}\beta w_j,\\
& D_j T_j + w_j = \beta T_j,\\
& \tau D_j S_j + w_j = \beta S_j, \\
& ij\alpha u_j + w_j'=0,\\
& u_j' \mid_{z=0,1} = w_j\mid_{z=0,1}= T_j\mid_{z=0,1}= S_j\mid_{z=0,1}=0
\end{aligned} 
\right.
\ee
for $j\in \mathbb Z$, where $'=d/dz$, $D_j=d^2/dz^2-j^2\alpha^2$. We reduce 
(\ref{eq4.3})  to a single equation for $w_j(z)$:
\begin{align}
\label{eq4.4}
& \{(\tau D_j-\beta)(D_j-\beta)(D_j-\sigma^{-1}\beta)D_j 
\\
& \qquad + j^2\alpha^2[\lambda(\tau D_j-\beta)-\eta(D_j-\beta)]\}w_j=0, 
\nonumber \\
& \label{eq4.5}
 w_j = w_j'' = w_j^{(4)} = w_j^{(6)} = 0  \quad \text{at} \quad z=0,1,
\end{align}
for $j\in \mathbb Z$. Thanks to ({\ref{eq4.5}}), $w_j$
can be expanded in a  Fourier sine series
\be
\label{eq4.6}
w_j(z)=\sum_{l=1}^{\infty}w_{jl}\sin l\pi z
\ee
for $j\in \mathbb Z$. Substituting (\ref{eq4.6}) into (\ref{eq4.4}),
we see that the eigenvalues $\beta$ of the problem (\ref{eq4.1}) satisfy
the following cubic equations
\begin{align}
& \label{eq4.7}
\beta^3+(\sigma+\tau+1)\gamma_{jl}^2\beta^2+[(\sigma+\tau+\sigma \tau)
\gamma_{jl}^4-\sigma j^2\alpha^2\gamma_{jl}^{-2}(\lambda-\eta)]\beta
+\sigma \tau \gamma_{jl}^6 \\
& \qquad \qquad \qquad + \sigma j^2 \alpha^2 (\eta - \tau \lambda)
=0, \nonumber 
\end{align}
for $j\in \mathbb Z$ and $l\in \mathbb N$,
 where $\gamma_{jl}^2=j^2 \alpha^2 + l^2\pi^2$.

For the sake of convenience to analyze the distribution of the eigenvalues, 
we  make the following  definitions.

\begin{definition}
\label{df4.1}
For fixed parameters $\sigma$, $\tau$, $\eta$ and $\lambda$, let
\begin{enumerate}

\item  $g_{jl}(\beta)=\beta^3+(\sigma+\tau+1)\gamma_{jl}^2\beta^2+
(\sigma+\tau+\sigma \tau)\gamma_{jl}^4\beta+\sigma \tau \gamma_{jl}^6$,\\

\item  $h_{jl}(\beta)=[\sigma j^2\alpha^2\gamma_{jl}^{-2}(\lambda-\eta)]\beta
-\sigma j^2 \alpha^2 (\eta - \tau \lambda)$,\\

\item  $f_{jl}(\beta)=g_{jl}(\beta)-h_{jl}(\beta)$,\\

\item $\eta_{c}=\frac{27}{4}\pi^4\tau^2(1+\sigma^{-1})(1-\tau)^{-1}$,\\

\item $\eta_{c_1}=\frac{27}{4}\pi^4\tau^3(1-\tau^2)^{-1},$\\

\item $\lambda_{c}=\frac{\eta}{\tau}+\frac{27}{4}\pi^4$,\\

\item  $\beta_{jl1},\beta_{jl2}$ and $\beta_{jl3}$ be the zeros of 
$f_{jl}$ with $ Re(\beta_{jl1})\geq Re(\beta_{jl2}) 
\geq Re(\beta_{jl3})$.

\end{enumerate}
\end{definition}

In the following discussions, we shall focus on the following 
diffusive regime:
\be
\label{eq4.8}
\quad \sigma>1>\tau>0,\,\, \alpha^2=\pi^2/2,\,\, \eta<\eta_{c}\, 
\text{and}\,\, \lambda\approx \lambda_c. 
\ee

\begin{lemma}
\label{le4.2}
\begin{enumerate}
\item Under the assumption (\ref{eq4.8}), $f_{11}(\beta)$ has three 
      simple real zeros.
\item If $\tau>1$, $\beta_{111}=0$ is a simple zero of $f_{11}(\beta)$
      for $\lambda=\lambda_c$.
\end{enumerate}
\end{lemma} 
\begin{proof}
Since $\lambda \approx \lambda_c$, it suffices to prove this statement for
 $\lambda=\lambda_c$. In this case,
\begin{align*}
 f_{11}(\beta)=& \beta^3+[3\pi^2(\sigma + \tau +1)/2]\beta^2
 +[9\pi^4(\sigma + \tau + \sigma \tau)/4 - \sigma(\lambda_c -\eta)/3] \beta\\
              =& f(\beta) \beta.
\end{align*}
Since $\eta < \eta_c$ or $\tau>1$, the constant term of $f(\beta)$ is nonzero.
Hence $\beta_{111}=0$ is a simple zero of $f_{11}$.
Moreover, for condition (\ref{eq4.8}), the quadratic discriminant of 
$f(\beta)$ is $9\pi^4 (\sigma + 1 - \tau)^2/4 + 
4 \sigma \eta (1- \tau)/(3 \tau) >0 $ . 
This implies $f_{11}$ has three 
simple real zeros.
\end{proof}

 We summarize the following important lemma about the distribution of 
 the zeros of $f_{jl}$.

\begin{lemma}
\label{le4.3}
 Assume that either
\begin{enumerate}
\item[1)]  $\eta<\eta_c$ with condition (\ref{eq4.8}) or
\item[2)]    $\eta>0$ with $\tau>1$,
\end{enumerate}
then
\begin{align}
& \label{eq4.9}
\beta_{111}(\lambda)
\left\{
\begin{aligned}
& <0 &&\qquad  \text{if} \,\,\,\, \lambda<\lambda_c,\\
& =0 &&\qquad  \text{if} \,\,\,\, \lambda=\lambda_c,\\
& >0 &&\qquad  \text{if} \,\,\,\, \lambda>\lambda_c,
\end{aligned}\right. 
\\
& \label{eq4.10}
Re\beta_{jlk}(\lambda)<0 && \text{for}\,\,\,  (j,l,k)\ne(1,1,1).
\end{align}           
\end{lemma}

\begin{proof}
Let
\begin{align*} 
f(\beta)=\beta^3+a_2\beta^2+a_1\beta+a_0
\end{align*}
be a monic real coefficient polynomial of degree 3. The following cases 
are apparent.  
\begin{enumerate}
\item  $\beta=0$ is a zero of $f(\beta)$ if and only if $a_0=0$.
\item  $\beta=bi,-bi,a$ ($a,b\in \mathbb R$) are  zeros 
       of $f(\beta)$ if and only if $a_1=b^2$, $a_2=-a$ and $a_0=-ab^2$.
\end{enumerate}
Case (2) is equivalent to
\begin{align*}
       a_1>0 \,\,\text{and}\,\, a_1a_2=a_0.
\end{align*}           

By the above observation, we prove the lemma in several steps 
as follows.

\medskip

{\sc Step 1.}
It's easy to see that $\beta=-\gamma_{0l}^2$, $-\sigma\gamma_{0l}^2$,
$-\tau\gamma_{0l}^2$ are the zeros of $f_{0l}$. 
When $j\ne 0$,  $\beta=0$  is a zero of $f_{jl}$  if and only if 
$$
\sigma\tau\gamma_{jl}^6+\sigma j^2 \alpha^2(\eta-\tau \lambda)=0, $$
which is equivalent to 
\begin{equation}
\label{eq4.11}
\lambda=\frac{\eta}{\tau}+\frac{\gamma_{jl}^6}{j^2\alpha^2}.
\end{equation}
For fixed $\eta$, minimizing the right hand side of (\ref{eq4.11}), we 
obtain that $\lambda_c=\frac{\eta}{\tau}+\frac{27}{4}\pi^4$ is the 
global minimum  of $\lambda$ in Case (1),
 provided $\alpha^2=\frac{\pi^2}{2}$ and $(j,l)=(1,1)$.

\medskip

{\sc Step 2.}  For Case (2), we obtain in the same fashion as above that 
\begin{equation}
\label{eq4.12}
\lambda=\frac{(\sigma+\tau)}{(\sigma+1)}\eta+
\frac{\gamma_{jl}^6}{\sigma j^2 \alpha^2}(\sigma+\tau)(\tau+1).  
\end{equation}
For a fixed  $\eta$, minimizing the right hand side of (\ref{eq4.12}), 
we obtain that
\begin{align*} 
\lambda_{c_1}=\frac{(\sigma+\tau)}{(\sigma+1)}\eta+
      \frac{27}{4}\pi^4(1+\sigma^{-1}\tau)(1+\tau) 
\end{align*}
is the global minimum
of $\lambda$ in the case (2), provided 
$\alpha^2=\frac{\pi^2}{2}$ and $(j,l)=(1,1)$.

\medskip

{\sc Step 3.} As introduced before, 
$\lambda=\lambda_c(\eta)$ and $\lambda=\lambda_{c_1}(\eta)$ 
define two straight lines in the $\lambda-\eta$ plane, shown in 
Figures~\ref{fig4.1}  and \ref{fig4.2}.

If $\tau<1$, the intersection of the two lines is
\begin{align*}
(\lambda_{c_2} ,\eta_c)=(\frac{27}{4}\pi^4\tau(1+\tau\sigma^{-1})(1-\tau), \frac{27}{4}\pi^4\tau^2(1+\sigma^{-1})(1-\tau)^{-1}).
\end{align*}
As shown in Figure~\ref{fig4.1}, 
$\lambda_c(\eta)<\lambda_{c_1}(\eta)$ for $\eta<\eta_c$.
If $\tau>1$, $\lambda_c(\eta)<\lambda_{c_1}(\eta)$ for $\eta>0$; see 
Figure~\ref{fig4.2}.
It is easy then to see that  in either case, (\ref{eq4.9}) and (\ref{eq4.10}) 
hold true. The proof is complete.
\end{proof}

\begin{remark}
\label{rm4.4}
{\rm  \begin{enumerate}
\item This lemma  works for the 3D double-diffusive problem as well.
\item In case 1), if $\eta>\eta_c$ then $\beta_{111}=\bar{\beta}_{112}$
      are complex numbers for $\lambda\approx \lambda_{c_1}$.
\item The distribution of the zeros of $f_{jl}$ was first 
      analyzed by Veronis \cite{gv}. The results are scattered
      in  different papers. To make this paper more self-contained,
      the authors think it's good to summarize it here 
      and give a clear proof.
\end{enumerate}
}
\end{remark}

 To check that the operators $-L_{\lambda \eta}$ 
satisfy condition (\ref{eq3.4}), we prove the following lemma.
\begin{lemma}
\label{le4.5}
\begin{enumerate}
\item Only finitely many numbers of the zeros of  $f_{jl}(\beta)$ have nonzero 
imaginary parts for $(j,l)\in \mathbb Z \times \mathbb N$.
\item $\beta_{jlk}\to-\infty$  \quad as  $j^2+l^2\to \infty$.
\end{enumerate}
\end{lemma}

\begin{proof}
Since $f_{jl}=g_{jl}-h_{jl}$, $\beta$ is a zero of $f_{jl}(\beta)$ if and only
 if $\beta$ satisfies the equation
\begin{equation}
\label{eq4.13} 
g_{jl}(\beta)=h_{jl}(\beta).
\end{equation}
Plugging  $\beta=\gamma_{jl}^{2}\beta^{*}$ into (\ref{eq4.13}), we obtain
\begin{equation}
\label{eq4.14}
(\beta^{*}+1)(\beta^{*}+\tau)(\beta^{*}+\sigma)=\vartheta_{jl}[(\lambda-\eta)
\beta^{*}-(\eta-\tau\lambda)],
\end{equation} 
where $\vartheta_{jl}=j^{2}\alpha^2\sigma/\gamma_{jl}^{6}$. Since 
$\displaystyle \lim_{j^2+l^2\to\infty}\vartheta_{jl}=0$, 
the roots of (\ref{eq4.14}) must be  negative 
real numbers near the interval $[-\sigma, -\tau]$ when $(j^2+l^2)$ is large.
This completes the proof. 
\end{proof}
 
\subsection{Eigenvectors}

 Let's make some observations to analyze the spectrum. 
Since $g_{jl}(\beta)=(\beta+\gamma_{jl}^2)
(\beta+\tau\gamma_{jl}^2)(\beta+\sigma\gamma_{jl}^2)$ and $h_{jl}=
\sigma j^2\alpha^2\gamma_{jl}^{-2}[(\lambda-\eta)\beta
-(\eta - \tau \lambda)\gamma_{jl}^2]$, it's easy to check that 
$\beta=-\gamma_{jl}^2$ or $\beta=-\tau\gamma_{jl}^2$ is a zero 
of $f_{jl}(\beta)$ if and only if $j=0$. In the case of $j=0$, the
zeros of $f_{jl}$ are $-\gamma_{jl}^2$, $-\tau\gamma_{jl}^2$ and
$-\sigma\gamma_{jl}^2$. The corresponding eigenvectors are 
\begin{align}
\label{eq4.15}
& \psi_{0l}^{1}(x,z)=( 0, 0 , \sin l\pi z , 0)^t,\\      
& \psi_{0l}^{2}(x,z)=(0, 0, 0, \sin l\pi z)^t,
\nonumber\\ 
& \psi_{0l}^{3}(x,z)=(\cos l\pi z, 0, 0, 0)^t.
\nonumber                 
\end{align}
To analyze the structures of the eigenspaces 
of problem (\ref{eq4.1}), we make the following definitions.
\begin{definition}
\label{df4.7}
For $j\ne 0$, we define
\begin{align*}
& \phi_{jl}^1(x,z)=( 
            \frac{l\pi}{j\alpha}\cos j\alpha x \cos l\pi z,
            \sin j\alpha x \sin l\pi z,
            0,
            0)^t,\\
&\phi_{jl}^2(x,z)=(
            0,
            0,
            \sin j\alpha x \sin l\pi z,
            0)^t, \,\,
\phi_{jl}^3(x,z)=( 
            0,
            0,
            0,
            \sin j\alpha x \sin l\pi z)^t,\\
&\phi_{jl}^4(x,z)=(
            -\frac{l\pi}{j\alpha}\sin j\alpha x \cos l\pi z,
            \cos j\alpha x \sin l\pi z,
            0,
            0)^t,\\
&\phi_{jl}^5(x,z)=(
           0,
           0,
           \cos j\alpha x \sin l\pi z,
           0)^t, \,\,
\phi_{jl}^6(x,z)=( 
            0,
            0,
            0,
            \cos j\alpha x \sin l\pi z)^t.
\end{align*}
for each $l\in \mathbb N$.
\end{definition}

\begin{lemma}
\label{le4.6}
If $j\ne 0$ and  $\beta$  
is a zero of $f_{jl}$, then we have the followings. 
\begin{enumerate}
\item  The eigenvector corresponding 
to $\beta$ in the complexified space is 
\begin{equation}
\label{eq4.16}
\psi_{jl}^{\beta}(x,z)=e^{ij\alpha x}( 
            \frac{il\pi}{j\alpha}\cos l\pi z 
           ,\sin l\pi z 
           , A_1(\beta)\sin l\pi z 
           , A_2(\beta)\sin l\pi z)^t, 
\end{equation} 
\begin{eqnarray*}
 \text{where} \qquad
A_1(\beta)=\frac{1}{\beta+\gamma_{jl}^2} \qquad \text{and} \qquad
 A_2(\beta)=\frac{1}{\beta+\tau\gamma_{jl}^2}.
\end{eqnarray*}
\item If $\beta$ is a real number, the corresponding  eigenvectors are
      given by
\begin{align}
\label{eq4.17}
&\psi_{jl}^{\beta,1}=\phi_{jl}^1+ A_1(\beta)\phi_{jl}^2+
                A_2(\beta)\phi_{jl}^3
               \qquad \text{and}\\
&\psi_{jl}^{\beta,2}=\phi_{jl}^4+ A_1(\beta)\phi_{jl}^5+ 
             A_2(\beta)\phi_{jl}^6.
\nonumber
\end{align} 
\item If $Im(\beta)\ne 0$, the generalized eigenvectors corresponding
      to $\beta$ and $\bar{\beta}$ are       
\begin{equation}
\label{eq4.18}
\begin {aligned}
& \psi_{jl}^{\beta,1}=\phi_{jl}^1+
             R_1(\beta)\phi_{jl}^2+
             R_2(\beta)\phi_{jl}^3
             +I_1(\beta)\phi_{jl}^5  
            +I_2(\beta)\phi_{jl}^6,\\
& \psi_{jl}^{\beta,2}=-I_1(\beta)\phi_{jl}^2
                      -I_2(\beta)\phi_{jl}^3
                      +\phi_{jl}^4
                      +R_1(\beta)\phi_{jl}^5
                      +R_2(\beta)\phi_{jl}^6,\\
& \psi_{jl}^{\bar{\beta},1}=\phi_{jl}^1+
             R_1(\bar{\beta})\phi_{jl}^2+
             R_2(\bar{\beta})\phi_{jl}^3
             +I_1(\bar{\beta})\phi_{jl}^5  
            +I_2(\bar{\beta})\phi_{jl}^6 \,\,\, \text{and}\\
& \psi_{jl}^{\bar{\beta},2}=-I_1(\bar{\beta})\phi_{jl}^2
                      -I_2(\bar{\beta})\phi_{jl}^3
                      +\phi_{jl}^4
                      +R_1(\bar{\beta})\phi_{jl}^5
                      +R_2(\bar{\beta})\phi_{jl}^6,\\
\end{aligned}
\end{equation}
where $R_1(\beta)=Re(A_1(\beta))$, $I_1(\beta)=Im(A_1(\beta))$, $R_2(\beta)=
Re(A_2(\beta))$ and $I_2(\beta)=Im(A_2(\beta))$.
\end{enumerate}
\end{lemma}
 
The proof of  Lemma~\ref{le4.6} follows from a direct calculation, 
and we shall omit the details. 

\begin{definition}
\label{df4.8}
\begin{enumerate}
\item If $j=0$  and $l\in \mathbb N$, we define 
$$E_{0l}=\text{span}\{\psi_{0l}^{1}(x,z),
                \psi_{0l}^{2}(x,z),
                \psi_{0l}^{3}(x,z)\}.$$ 
\item For $j\in \mathbb N$, we define 
        $E_{jl}^1=$ span $\{\phi_{jl}^1(x,z),\phi_{jl}^2(x,z),\phi_{jl}^3(x,z)\}$, 
        $E_{jl}^2=$ span $\{\phi_{jl}^4(x,z),\phi_{jl}^5(x,z),\phi_{jl}^6(x,z)\}$
      and $E_{jl}=E_{jl}^1\oplus E_{jl}^2$.

\item For $ j\in{0}\cup\mathbb N$, $l\in \mathbb N$, we define $E_{f_{jl}}$
      be the eigenspace spanned by the eigenvectors and the generalized 
      eigenvectors corresponding to the zeros of $f_{jl}$. 
\end{enumerate}
\end{definition}

It is easy to see that the completion of $\oplus_{j=0,l=1}^{\infty}E_{jl}$
 in H-norm is H. Hence the following theorem shows that the eigenvectors 
 and the generalized eigenvectors corresponding to the zeros of 
  $\{f_{jl}\}_{j=0,l=1}^{\infty}$ form a basis of H.

\begin{theorem}
\label{th4.9}
Under the assumption (\ref{eq4.8}), we have
\begin{enumerate}
\item[1)] $E_{f_{jl}}=E_{jl}$ for $j\in\{0\}\cup \mathbb N$, $l\in \mathbb N$;
      and
\item[2)] $L_{\lambda \mu}|E_{jl}$ is strictly negative definite for each
      $(j,l)\in \mathbb Z \times \mathbb N$ when $\lambda < \lambda_c$. 

\end{enumerate}
\end{theorem}
\begin{proof} We proceed in two steps.

\medskip

{\sc Step 1.}
To prove Assertion 1), it is enough to show that 
dim $E_{f_{jl}}\ge$ dim $E_{jl}$.
The case of $j=0$ follows from (\ref{eq4.15}).
When $j\ne 0$, for  a fixed $l$, we examine all the cases as follows.

1. If $\beta_1>\beta_2>\beta_3$ are distinct zeros of $f_{jl}(\beta)$,
      by (\ref{eq4.17}), we have dim $E_{f_{jl}}\ge 6=$ dim $E_{jl}$ .

\medskip

2. If $\beta_1=\bar{\beta_2} \in \mathbb C\backslash \mathbb R $ 
      and $\beta_3$
      are distinct zeros of $f_{jl}(\beta)$,  
      by  (\ref{eq4.17}) and (\ref{eq4.18}),
      we have dim $E_{f_{jl}}\ge 6=$ dim $E_{jl}$ .

\medskip

3. The first derivative of $f_{jl}$ is $f_{jl}'=3\beta^2+
      2(\sigma+\tau+1)\gamma_{jl}^2\beta+
      [(\sigma+\tau+\sigma\tau)\gamma_{jl}^4-
      \sigma j^2\alpha^2\gamma_{jl}^{-2}(\lambda-\eta)]$. The quadratic
      discriminant of $f_{jl}'$ is 
      \[
      4\{\frac{1}{2}[(\sigma-\tau)^2+(\sigma-1)^2+(\tau-1)^2]\gamma_{jl}^4
      + 3\sigma j^2\alpha^2\gamma_{jl}^{-2}(\lambda-\eta)\}>0,
      \]
      since $\lambda\approx \lambda_c>\eta$. It follows that $f_{jl}(\beta)$
      cannot have  zeros of multiplicity 3.

\medskip

4. If $\beta_1\ne \beta_2=\beta_3$ are zeros of $f_{jl}$, direct 
      computation shows
      \begin{align}
      L_{\lambda \eta}\phi_{jl}^2=
           & \frac{(\sigma\lambda j^2\alpha^2)}{\gamma_{jl}^2}\frac{(\beta_1+
      \tau\gamma_{jl}^2)}{(\beta_1-\beta_2)}\psi_{jl}^{\beta_1,1}
      \label{eq4.19} \\
           & +\frac{(\sigma \lambda j^2\alpha^2)}{\gamma_{jl}^2}\frac{(\beta_2+
      \tau\gamma_{jl}^2)}{(\beta_2-\beta_1)}\psi_{jl}^{\beta_2,1}
       \nonumber  \\
           & +[\frac{(\sigma\lambda j^2\alpha^2)(\tau-1)}{(\beta_1+
         \gamma_{jl}^2)(\beta_2+\gamma_{jl}^2)} - \gamma_{jl}^2] \phi_{jl}^2
       \nonumber  
       \end{align}

       Note that
       \begin{align*}
        (\beta_1+ & \gamma_{jl}^2)(\beta_2+\gamma_{jl}^2)(\beta_3+\gamma_{jl}^2)
         =-f_{jl}(-\gamma_{jl}^2)\\
       &=h_{jl}(-\gamma_{jl}^2)-g_{jl}(-\gamma_{jl}^2)
         = h_{jl}(-\gamma_{jl}^2)\\
       &=\sigma j^2 \alpha^2 \lambda (\tau-1).
       \end{align*}
       Hence 
       \begin{eqnarray*}
       \beta_3=\frac{(\sigma\lambda j^2\alpha^2)(\tau-1)}{(\beta_1+
         \gamma_{jl}^2)(\beta_2+\gamma_{jl}^2)} - \gamma_{jl}^2.
       \end{eqnarray*}
       We pick up $v=m\phi_{jl}^2$ to be
       the generalized eigenvector corresponding to $\beta_3$ in 
       $E_{jl}^{1}$,  where $m$ is some small constant. It's easy to check
       that  $E_{jl}^{1}=$ span$\{ \psi_{jl}^{\beta_1,1},\psi_{jl}^{\beta_2,1},
       v \}$ and  $L_{\lambda \eta}|E_{jl}^1$ can be represented by matrix
       \begin{equation}
       \label{eq4.20} 
       \left(\begin{array}{ccc}
        \beta_1  & 0  & m \frac{(\sigma\lambda j^2\alpha^2)}{\gamma_{jl}^2}
            \frac{(\beta_1+\tau\gamma_{jl}^2)}{(\beta_1-\beta_2)}  \\
        0  & \beta_2  & m \frac{(\sigma \lambda j^2\alpha^2)}{\gamma_{jl}^2}
           \frac{(\beta_2+\tau\gamma_{jl}^2)}{(\beta_2-\beta_1)}   \\
        0 & 0 & \beta_3 
       \end{array}
       \right)
       \end{equation}
       in the  basis $\{\psi_{jl}^{\beta_1,1},\psi_{jl}^{\beta_2,1},
       v \}$.
       The same argument works for $E_{jl}^2$ as well.    
 
\medskip

{\sc Step 2.}
It's easy to check that $E_{jl}^1$ is orthogonal to 
$E_{jl}^2$ for $(j,l)\in \mathbb N \times \mathbb N$ and 
$E_{j_1l_1}$ is orthogonal to $E_{j_2l_2}$ for 
$(j_1,l_1) \ne (j_2,l_2)$. Lemma~\ref{le4.3} together with
Step1 imply that  $L_{\lambda \eta}|E_{jl}$ is strictly
negative definite when $\lambda <\lambda_c$.
This completes the proof.
\end{proof}
Lemma~\ref{le4.3} and Lemma~\ref{le4.5} together with Theorem~\ref{th4.9} 
imply the following theorem. 
\begin{theorem}
\label{th4.10}
Under assumption (\ref{eq4.8}), $-L_{\lambda \eta}$ is a sectorial operator.
\end{theorem}
\begin{remark}
\label{rm4.11}
{\rm
\begin{enumerate}
\item Since dim $E_{jl}=3$ or $6$ which is finite, there exists
      a vector $\Psi_{jl}^{\beta,k} \in E_{jl}$ such that
      \begin{eqnarray*} 
 <\Psi_{jl}^{\beta,k}, \psi_{jl}^{\beta^*,k^*}>_H\begin{cases}
      &=0 \quad \text{for}\,\, (\beta ,k)\ne (\beta^*,k^*),\\
      &\ne 0 \quad \text{for}\,\, (\beta ,k)= (\beta^*,k^*),
     \end{cases} 
     \end{eqnarray*}
      where $\beta$ and 
      $\beta^*$ are zeros of $f_{jl}$ and $k,k^*=1,2$.

\item Note that $E_{jl}^1$ is orthogonal to 
      $E_{jl}^2$ for $(j,l)\in \mathbb N \times \mathbb N$ and 
      $E_{j_1l_1}$ is orthogonal to $E_{j_2l_2}$ for 
      $(j_1,l_1) \ne (j_2,l_2)$, hence we conclude that   
      $<\Psi_{j_1l_1}^{\beta,k_1},\psi_{j_2l_2}^{\beta^*, k_2}>_H=0$ 
      for $(j_1,l_1,\beta,k_1)\ne (j_2,l_2,\beta^*,k_2)$.

\item  If $j=0$, we  pick up 
$\Psi_{0l}^k=\psi_{0l}^k$. For $(j,l)=(1,1)$, we  pick up
\begin{equation}
\label{eq4.21}
 \Psi_{11}^{\beta_{111},1}= \phi_{11}^1 + C_1 \phi_{11}^2 
   + C_2 \phi_{11}^3, \qquad \text{and} \qquad 
 \Psi_{11}^{\beta_{111},2}= \phi_{11}^4 +
 C_1 \phi_{11}^5 + C_2 \phi_{11}^6,
\end{equation}
 where 
\begin{equation}
\label{eq4.22}
C_1=\frac{\sigma \lambda}{\beta_{111}+\gamma_{11}^2}
\qquad \text{and} \qquad
 C_2=\frac{-\sigma \eta}{\beta_{111}+\tau \gamma_{11}^2}.
\end{equation}

\item Lemma~\ref{le4.2} and Theorem~\ref{th4.9} show that  
the multiplicity of the eigenvalue  $\beta_{111}(\lambda)$ is two 
and the corresponding eigenvectors are $\psi^{\beta_{111},1}_{11}$
 and $\psi^{\beta_{111},2}_{11}$.

\item  For $\psi\in H_{3/4}\subset H$, by Sobolev inequality, 
 \begin{eqnarray*}
  |G(\psi)|_{H}^2  \le\int_0^1\int_0^{2\pi/\alpha}|\psi|^2|\nabla\psi|^2dxdz                  \le |\psi|_{L^{\infty}}^2|\psi|_{H_{1/2}}^2
                 \le C |\psi|_{H_{3/4}}^4. 
 \end{eqnarray*}
   where $C$ is some constant. Hence, $G(\psi)=o(|\psi|_{H_{3/4}})$.
\end{enumerate}
}
\end{remark}

\section {Center manifold reduction}
We are now in a position to  reduce equations of (\ref{eq2.4})-
(\ref{eq2.7}) to the center manifold. We would like to fix 
$\eta <\eta_c$ , and let $\lambda \approx \lambda_c$ be the bifurcation 
parameter. For any $\psi=(U,T,S)\in H$, we have 
\begin{eqnarray*}
\psi=\sum_{j=0,l=1}^{\infty}\sum_{k=1}^{3}(x_{jlk}\psi_{jl}^{\beta_{jlk},1}+
     y_{jlk}\psi_{jl}^{\beta_{jlk},2}).  
\end{eqnarray*}
Since $\beta_{111}$ is the first eigenvalue, $\psi_{11}^{\beta_{111},1}$  and 
$\psi_{11}^{\beta_{111},2}$ are the first eigenvectors.
The reduced equations are given by

\begin{equation}
\label{eq5.1}
\left\{
\begin{aligned}
& \frac{dx_{111}}{dt} = \beta_{111}(\lambda) x_{111}+
   \frac{1}{<\psi_{11}^{\beta_{111},1},\Psi_{11}^{\beta_{111},1}>_H}
   <G(\psi,\psi),\Psi_{11}^{\beta_{111},1}>_H,\\
& \frac{dy_{111}}{dt}=\beta_{111}(\lambda)y_{111}+
   \frac{1}{<\psi_{11}^{\beta_{111},2},
   \Psi_{11}^{\beta_{111},2}>_H}<G(\psi,\psi),
    \Psi_{11}^{\beta_{111},2}>_H.
\end{aligned}
\right.
\end{equation}
Here for $\psi_1=(U_1,T_1,S_1)$, $\psi_2=(U_2,T_2,S_2)$ and 
$\psi_3=(U_3,T_3,S_3)$,
\begin{eqnarray*}
G(\psi_1,\psi_2)=-(  P (U_1\cdot\nabla)U_2,
                   (U_1\cdot\nabla)T_2,
                   (U_1\cdot\nabla)S_2
                   )^t \qquad \text{and}
\end{eqnarray*} 
\begin{eqnarray*} 
<G(\psi_1,\psi_2),\psi_3>_H=-\int_{\Omega}[<(U_1\cdot\nabla)U_2,
         U_3>_{\mathbb R^2}+(U_1\cdot\nabla)T_2 T_3 
        +(U_1\cdot\nabla)S_2 S_3]dxdz, 
\end{eqnarray*}
where P is the Leray projection to 
$L^2$ fields.

Let the center manifold function be denoted by 
\begin{equation}
\label{eq5.2}
\Phi=\sum_{\beta \ne \beta_{111}}( \Phi_{jl}^{\beta,1}(x_{111},
      y_{111})\psi_{jl}^{\beta,1}+
     \Phi_{jl}^{\beta,2}(x_{111},y_{111})\psi_{jl}^{\beta,2}).
\end{equation}

Note that for any $\psi_i \in H_1$($i=$ 1, 2, 3),
\begin{align*}
& <G(\psi_1,\psi_2),\psi_2>_H=0,\\
& <G(\psi_1,\psi_2),\psi_3>_H=-<G(\psi_1,\psi_3),\psi_2>_H;
\end{align*}
and for k=1,2,
\begin{eqnarray*}
 <G(\psi_1,\psi_{11}^{\beta_{111},k}),\Psi_{11}^{\beta_{111},k}>_H=0.
\end{eqnarray*}

Then by $\psi=x_{111}\psi_{11}^{\beta_{111}
        ,1}+y_{111}\psi_{11}^{\beta_{111},2}+\Phi$, we have

\begin{align}
 <G (\psi,\psi),\Psi_{11}^{\beta_{111},1}>_H=&
           <G(\psi_{11}^{\beta_{111},2},\psi_{11}^{\beta_{111}
            ,2}),\Psi_{11}^{\beta_{111},1}>_H y_{111}^2 
          \label{eq5.3}\\
    &+<G(\psi_{11}^{\beta_{111},1},\psi_{11}^{\beta_{111}
            ,2}),\Psi_{11}^{\beta_{111},1}>_H  x_{111}y_{111}
          \nonumber \\
    &-<G(\psi_{11}^{\beta_{111},1},
            \Psi_{11}^{\beta_{111},1}),\Phi>_H x_{111}
          \nonumber\\
    &-<G(\psi_{11}^{\beta_{111},2},
            \Psi_{11}^{\beta_{111},1}),\Phi>_H y_{111}
           \nonumber\\
    &+<G(\Phi,\psi_{11}^{\beta_{111}
             ,2}),\Psi_{11}^{\beta_{111},1}>_H y_{111}
          \nonumber\\  
    &+<G(\Phi,\Phi),\Psi_{11}^{\beta_{111},1}>_H,
          \nonumber   
\end{align}

\begin{align}
 <G(\psi,\psi),\Psi_{11}^{\beta_{111},2}>_H=
           &<G(\psi_{11}^{\beta_{111},1},\psi_{11}^{\beta_{111}
            ,1}),\Psi_{11}^{\beta_{111},2}>_H x_{111}^2 
          \label{eq5.4}\\
           &+<G(\psi_{11}^{\beta_{111},2},\psi_{11}^{\beta_{111}
            ,1}),\Psi_{11}^{\beta_{111},2}>_H  x_{111}y_{111}
          \nonumber \\
           &-<G(\psi_{11}^{\beta_{111},2},
            \Psi_{11}^{\beta_{111},2}),\Phi>_H y_{111}
          \nonumber\\
           &-<G(\psi_{11}^{\beta_{111},1},
            \Psi_{11}^{\beta_{111},2}),\Phi>_H x_{111}
          \nonumber\\
           &+<G(\Phi,\psi_{11}^{\beta_{111}
             ,1}),\Psi_{11}^{\beta_{111},2}>_H x_{111}
          \nonumber\\  
           &+<G(\Phi,\Phi),\Psi_{11}^{\beta_{111},1}>_H,
          \nonumber   
\end{align}

By direct calculations, we obtain that
\begin{equation}
\label{eq5.5}
\begin{aligned}
& G(\psi_{11}^{\beta_{111},1},\psi_{11}^{\beta_{111},1})=-
       (P(
       -\frac{\pi^2}{2\alpha} \sin 2 \alpha x,
       \frac{\pi}{2} \sin 2 \pi z), 
       \frac{A_1(\beta_{111}) \pi}{2} \sin 2 \pi z, 
       \frac{A_2(\beta_{111}) \pi}{2} \sin 2 \pi z
       )^t,\\
&  G(\psi_{11}^{\beta_{111},1},\psi_{11}^{\beta_{111},2})=-
       (
       P( \frac{-\pi^2}{2\alpha}(\cos 2\alpha x +\cos 2\pi z),
       0),
       0,
       0)^t,\\
&   G(\psi_{11}^{\beta_{111},2},\psi_{11}^{\beta_{111},1})=-
       (P(
       \frac{-\pi^2}{2\alpha}(\cos 2\alpha x -\cos 2\pi z),
       0),
       0,
       0 )^t,\\
& G(\psi_{11}^{\beta_{111},2},\psi_{11}^{\beta_{111},2})=-
       (P(
       \frac{\pi^2}{2\alpha} \sin 2 \alpha x,
       \frac{\pi}{2} \sin 2 \pi z),
       \frac{A_1(\beta_{111}) \pi}{2} \sin 2 \pi z,
       \frac{A_2(\beta_{111}) \pi}{2} \sin 2 \pi z)^t,
\end{aligned}
\end{equation}
 and
\begin{equation}
\label{eq5.6}
\begin{aligned}
& G(\psi_{11}^{\beta_{111},1},\Psi_{11}^{\beta_{111},1})=-
       (P(  -\frac{\pi^2}{2\alpha} \sin 2 \alpha x,
       \frac{\pi}{2} \sin 2 \pi z ),
       \frac{C_1(\beta_{111}) \pi}{2} \sin 2 \pi z ,
       \frac{C_2(\beta_{111}) \pi}{2} \sin 2 \pi z)^t,\\
&  G(\psi_{11}^{\beta_{111},1},\Psi_{11}^{\beta_{111},2})=-
       (P( \frac{-\pi^2}{2\alpha}(\cos 2\alpha x +\cos 2\pi z),
       0),
       0,
       0)^t,\\
&   G(\psi_{11}^{\beta_{111},2},\Psi_{11}^{\beta_{111},1})=-
       ( P(\frac{-\pi^2}{2\alpha}(\cos 2\alpha x -\cos 2\pi z),
       0),
       0,
       0)^t,\\
& G(\psi_{11}^{\beta_{111},2},\Psi_{11}^{\beta_{111},2})=-
       (P(\frac{\pi^2}{2\alpha} \sin 2 \alpha x,
       \frac{\pi}{2} \sin 2 \pi z),
       \frac{C_1(\beta_{111}) \pi}{2} \sin 2 \pi z,
       \frac{C_2(\beta_{111}) \pi}{2} \sin 2 \pi z )^t.
\end{aligned}
\end{equation}

By (\ref{eq5.5}) and (\ref{eq5.6}),  we derive that
 for $(j,l)\ne(0,2)$,
\begin{equation}
\label{eq5.7}
\begin{aligned}
& <G(\psi_{11}^{\beta_{111},k_1},
  \psi_{11}^{\beta_{111},k_2}),\Psi_{jl}^{\beta,k}>_H=0,\\
& <G(\psi_{jl}^{\beta,k},
  \psi_{11}^{\beta_{111},k_1}),\Psi_{11}^{\beta_{111},k_2}>_H=0,
\end{aligned}
\end{equation}
where $k_1, k_2 =1,2.$
Hence the first two terms in (\ref{eq5.3}) and (\ref{eq5.4}) are gone.
Since the center manifold function contains only higher order terms
\begin{eqnarray*}
 \Phi(x_{111},y_{111})=O(|x_{111}|^2,|y_{111}|^2),
\end{eqnarray*}
we derive that 
\begin{equation}
\label{eq5.8}
\left\{
\begin{aligned}
& <G(\Phi,\Phi),\Psi_{11}^{\beta_{111},1}>_H=o(|x_{111}|^3,|y_{111}|^3),\\
& <G(\Phi,\Phi),\Psi_{11}^{\beta_{111},2}>_H=o(|x_{111}|^3,|y_{111}|^3).
\end{aligned}
\right.
\end{equation}

By (\ref{eq5.7}) and (\ref{eq5.8}), only 
$\Phi_{02}^{\beta_1}(x_{111},y_{111})$, 
$\Phi_{02}^{\beta_2}(x_{111},y_{111})$ and 
$\Phi_{02}^{\beta_3}(x_{111},y_{111})$ 
(where $\beta_1=-\gamma_{02}^2=-4 \pi^2$ , 
  $\beta_2=-\tau \gamma_{02}^2=-4 \tau \pi^2$, and
  $ \beta_3=-\sigma \gamma_{02}^2=-4 \sigma \pi^2$. ) contribute
to the third order terms in evaluation of (\ref{eq5.3}) and (\ref{eq5.4}).
Direct calculations show

\begin{align*}
&  <G(\psi_{11}^{\beta_{111},1},\psi_{11}^{\beta_{111},1}),
  \Psi_{02}^{\beta_1}>_H=\int_0^1 \int_0^{\frac{2\pi}
  {\alpha}}-\frac {A_1(\beta_{111})\pi}{2} \sin^2 2\pi z dxdz
  =\frac{-A_1(\beta_{111})\pi^2}{2\alpha},\\
&  <G(\psi_{11}^{\beta_{111},2},\psi_{11}^{\beta_{111},2}),
  \Psi_{02}^{\beta_1}>_H=\int_0^1 \int_0^{\frac{2\pi}
 {\alpha}}-\frac {A_1(\beta_{111})\pi}{2} \sin^2 2\pi z dxdz
 =\frac{-A_1(\beta_{111})\pi^2}{2\alpha},\\
& <G(\psi_{11}^{\beta_{111},2},\psi_{11}^{\beta_{111},1}),\Psi_{02}^{\beta_1}>_H=0, \qquad
  <G(\psi_{11}^{\beta_{111},1},\psi_{11}^{\beta_{111},2}),
  \Psi_{02}^{\beta_1}>_H=0,\\
&  <G(\psi_{11}^{\beta_{111},1},\psi_{11}^{\beta_{111},1}),
  \Psi_{02}^{\beta_2}>_H=\int_0^1 \int_0^{\frac{2\pi}
  {\alpha}}-\frac {A_2(\beta_{111})\pi}{2} \sin^2 2\pi z dxdz
  =\frac{-A_2(\beta_{111})\pi^2}{2\alpha},\\
&  <G(\psi_{11}^{\beta_{111},2},\psi_{11}^{\beta_{111},2}),
  \Psi_{02}^{\beta_2}>_H=\int_0^1 \int_0^{\frac{2\pi}
 {\alpha}}-\frac {A_2(\beta_{111})\pi}{2} \sin^2 2\pi z dxdz
 =\frac{-A_2(\beta_{111})\pi^2}{2\alpha},\\
& <G(\psi_{11}^{\beta_{111},2},\psi_{11}^{\beta_{111},1}),\Psi_{02}^{\beta_2}>_H=0, \qquad
  <G(\psi_{11}^{\beta_{111},1},\psi_{11}^{\beta_{111},2}),
  \Psi_{02}^{\beta_2}>_H=0,\\
&  <G(\psi_{11}^{\beta_{111},1},\psi_{11}^{\beta_{111},1}),
  \Psi_{02}^{\beta_3}>_H=0, \qquad
  <G(\psi_{11}^{\beta_{111},2},\psi_{11}^{\beta_{111},2}),
  \Psi_{02}^{\beta_3}>_H=0,\\
& <G(\psi_{11}^{\beta_{111},2},\psi_{11}^{\beta_{111},1}),\Psi_{02}^{\beta_3}>_H  = \frac{\pi^3}{2\alpha^2}, \qquad
  <G(\psi_{11}^{\beta_{111},1},\psi_{11}^{\beta_{111},2}),
  \Psi_{02}^{\beta_3}>_H=\frac{-\pi^3}{2\alpha^2},\\
& <\psi_{02}^{\beta_1},\Psi_{02}^{\beta_1}>_H=
<\psi_{02}^{\beta_2},\Psi_{02}^{\beta_2}>_H=
 <\psi_{02}^{\beta_3},\Psi_{02}^{\beta_3}>_H=\frac{\pi}{\alpha}.
\end{align*} 
 
Applying Theorem~\ref{th3.2}, we obtain
\begin{equation}
\label{eq5.9}
\begin{aligned}
& \Phi_{02}^{\beta_1}(x_{111},y_{111})=
       \frac{A_1(\beta_{111})\pi}{2\beta_1}(x_{111}^2+y_{111}^2)
       +o(x_{111}^2+y_{111}^2),\\
& \Phi_{02}^{\beta_2}(x_{111},y_{111})=
       \frac{A_2(\beta_{111})\pi}{2\beta_2}(x_{111}^2+y_{111}^2)
       +o(x_{111}^2+y_{111}^2),\\
& \Phi_{02}^{\beta_3}(x_{111},y_{111})=o(x_{111}^2+y_{111}^2). 
\end{aligned}
\end{equation}

By (\ref{eq5.5})-(\ref{eq5.9}), we evaluate

\begin{equation}
\label{eq5.10}
\left\{
\begin{aligned}
 & <G(\psi_{11}^{\beta_{111},1},\Psi_{11}^{\beta_{111},1}),\Phi>_H\\
  & = <G(\psi_{11}^{\beta_{111},1} ,\Psi_{11}^{\beta_{111},1}),
  \Phi_{02}^{\beta_1}\psi_{02}^{\beta_1}+\Phi_{02}^{\beta_2}
  \psi_{02}^{\beta_2}>_H + o(x_{111}^2+y_{111}^2)\\
  & = \frac{-C_1\pi^2}{2\alpha}\Phi_{02}^{\beta_1}-
   \frac{C_2\pi^2}{2\alpha}\Phi_{02}^{\beta_2}+o(x_{111}^2+y_{111}^2)\\
  & = -\frac{\pi^3}{4\alpha}(\frac{A_1C_1}{\beta_1}+\frac{A_2C_2}{\beta_2})
  (x_{111}^2+y_{111}^2)+o(x_{111}^2+y_{111}^2)\\
  & = \frac{1}{8\sqrt{2}}(A_1C_1+A_2C_2\tau^{-1})(x_{111}^2+y_{111}^2)+
  o(x_{111}^2+y_{111}^2),\\
& <G(\psi_{11}^{\beta_{111},2},\Psi_{11}^{\beta_{111},1}),\Phi>_H
  =o(x_{111}^2+y_{111}^2),\\
& <G(\Phi, \psi_{11}^{\beta_{111},2}),\Psi_{11}^{\beta_{111},1}>_H
  =o(x_{111}^2+y_{111}^2),\\
& <\psi_{11}^{\beta_{111},1},\Psi_{11}^{\beta_{111},1}>_H=\frac{1}{\sqrt{2}}
  (3+A_1C_1+A_2C_2),\\
\end{aligned}
\right.
\end{equation}
\begin{equation}
\label{eq5.11}
\left\{
\begin{aligned}
& <G(\psi_{11}^{\beta_{111},2}  ,\Psi_{11}^{\beta_{111},2}),\Phi>_H\\
& = <G(\psi_{11}^{\beta_{111},1} ,\Psi_{11}^{\beta_{111},2}),
  \Phi_{02}^{\beta_1}\psi_{02}^{\beta_1}+\Phi_{02}^{\beta_2}
  \psi_{02}^{\beta_2}>_H + o(x_{111}^2+y_{111}^2)\\
& = \frac{-C_1\pi^2}{2\alpha}\Phi_{02}^{\beta_1}-
   \frac{C_2\pi^2}{2\alpha}\Phi_{02}^{\beta_2}+o(x_{111}^2+y_{111}^2)\\
& = -\frac{\pi^3}{4\alpha}(\frac{A_1C_1}{\beta_1}+\frac{A_2C_2}{\beta_2})
  (x_{111}^2+y_{111}^2)+o(x_{111}^2+y_{111}^2)\\
& = \frac{1}{8\sqrt{2}}(A_1C_1+A_2C_2\tau^{-1})(x_{111}^2+y_{111}^2)+
  o(x_{111}^2+y_{111}^2),\\
& <G(\psi_{11}^{\beta_{111},1},\Psi_{11}^{\beta_{111},2}),\Phi>_H
  =o(x_{111}^2+y_{111}^2),\\
& <G(\Phi, \psi_{11}^{\beta_{111},1}),\Psi_{11}^{\beta_{111},2}>_H
  =o(x_{111}^2+y_{111}^2),\\
& <\psi_{11}^{\beta_{111},1},\Psi_{11}^{\beta_{111},1}>_H=\frac{1}{\sqrt{2}}
  (3+A_1C_1+A_2C_2),\\
\end{aligned}
\right.
\end{equation}

Plugging (\ref{eq5.10}) and (\ref{eq5.11}) into (\ref{eq5.3}) 
and (\ref{eq5.4}) respectively and applying Theorem~\ref{th3.2}, 
we get the reduced bifurcation equations: 
\begin{align}
 \frac{dx_{111}}{dt}=
         & \beta_{111}(\lambda)x_{111}
          -\frac{1}{8}\delta(\lambda, \eta)(x_{111}^3+x_{111}y_{111}^2)
       \label{eq5.12}\\
        & +o(x_{111}^3+y_{111}^3)
        +O(\beta_{111}(\lambda) (x_{111}^3+y_{111}^3) ),
        \nonumber
\end{align}
\begin{align}
 \frac{dy_{111}}{dt}=
         & \beta_{111}(\lambda)y_{111}
           -\frac{1}{8}\delta(\lambda,\eta)(x_{111}^2 y_{111}+y_{111}^3)
         \label{eq5.13}\\
         & +o(x_{111}^3+y_{111}^3)
           +O(\beta_{111}(\lambda)(x_{111}^3+y_{111}^3)),
        \nonumber
\end{align}
where 
\begin{equation}
\label{eq5.14}
\delta(\lambda, \eta)=\frac{(A_1C_1+A_2C_2 \tau^{-1})}{(3+A_1C_1+A_2C_2)}.
\end{equation}
 The following lemma determines the sign of $\delta(\lambda, \eta)$.
\begin{lemma}
\label{le5.1}
\begin{enumerate}
\item Under the assumption (\ref{eq4.8}),
\begin{align*}
\delta(\lambda, \eta)\left\{\begin{aligned}
&  >0  \quad  \text{if} \,\, \eta<\eta_{c_1},\\ 
&  <0  \quad  \text{if} \,\, \eta>\eta_{c_1}.
\end{aligned}\right.
\end{align*}
\item If we replace $\tau>1$ ($\tau\ne\sigma$) in (\ref{eq4.8}),
      then $\delta(\lambda,\eta)>0$ for all $\eta>0$.
\end{enumerate} 
\end{lemma}
\begin{proof}
{\sc Step 1.}
Under the assumption (\ref{eq4.8}), 
$\lambda \approx \lambda_c=\frac{\eta}{\tau}+\frac{27\pi^4}{4}$ 
and  $\beta_{111}\approx 0$. It suffices to prove the lemma for 
$\lambda=\frac{\eta}{\tau}+\frac{27\pi^4}{4}$ and  $\beta_{111}=0$.
Note that   
\begin{eqnarray*}
 A_1 C_1
   =\frac{\sigma \lambda}{(\beta_{111}+\gamma_{11}^2)^2}\,\,\text{and} \,\,
 A_2 C_2
 =\frac{-\sigma \eta}{(\beta_{111}+\tau \gamma_{11}^2)^2}.
\end{eqnarray*}
Plugging
  $\lambda=\frac{\eta}{\tau}+\frac{27\pi^4}{4}$ and 
$\beta_{111}=0$ into the denominator and numerator of $\delta(\lambda, \eta)$
respectively  yileds
\begin{align*}
3+A_1C_1+A_2C_2 & \approx 3+\gamma_{11}^{-4}(\sigma\frac{\eta}{\tau}
-\sigma\frac{\eta}{\tau^2})\\
& =\sigma \gamma_{11}^{-4}(1-\tau)\tau^{-2}(\eta_c-\eta)\\
&>0,    \\
\text{and} \qquad  A_1C_1+A_2C_2\tau^{-1} & \approx 
\sigma \gamma_{11}^{-4}(1-\tau^{2})\tau^{-3}(\eta_{c_1}-\eta).\\
\end{align*}
Hence the sign of $\delta(\lambda, \eta)$  is 
determined by $(\eta_{c_1}-\eta)$.

{\sc Step 2.}
If $\tau>1$, we have
\begin{align*}
& 3+A_1C_1+A_2C_2 \approx 3+\gamma_{11}^{-4}(\sigma\frac{\eta}{\tau}
-\sigma\frac{\eta}{\tau^2}) >0,\\
\text{and} \qquad &  A_1C_1+A_2C_2\tau^{-1} \approx 
\sigma \gamma_{11}^{-4}(\frac{27}{4}\pi^4-\eta(1-\tau^2)\tau^{-3})>0.
\end{align*}
Hence, $\delta(\lambda,\eta)>0$ for all $\eta>0$.
This completes the proof.
\end{proof}

\section{Completion of the proofs}
We demonstrate the proof of  Theorem~\ref{th2.3}; Theorem~\ref{th2.5}
can be proved in the same fashion.

\subsection{$S^1$-Attractor}

First, Assertion (1) of Theorem~\ref{th2.3}  follows 
from (\ref{eq5.12})-(\ref{eq5.14}) and Lemma~\ref{le5.1}. 
Then, by 
Theorem~\ref{th3.7}, the equations bifurcate from $(0,\lambda_c)$
to an attractor $\Sigma_{\lambda}$ for $\lambda>\lambda_c$, which is
homeomorphic to $S^1$. 

\subsection {Singularity Cycle}
We shall prove that the bifurcated attractor $\Sigma_{\lambda}$ of
(\ref{eq2.4})-(\ref{eq2.7}) given in Theorem~\ref{th2.3} 
is  a cycle of steady state solutions.
First, let 
\begin{align*}
& H'=\{(U,T,S)|u(-x,z)=-u(x,z)\},\\  
& H_{1}'=H_{1}\cap H'.
\end{align*}
It is well-known that $H'$ and $H'_{1}$ are invariant spaces for the operator
$L_{\lambda \eta}+G$ given by (\ref{eq2.11}) in the sense that 
\begin{align*}
L_{\lambda \eta}+G:H_1' \to H'. 
\end{align*}
It is clear that the first eigenvalue of $L_{\lambda \eta}|H_1'$ is simple
  when $\eta<\eta_c$. 
By the Kransnoselski  bifurcation theorem ( see aomng others Chow and Hale 
\cite{ch} and  Nirenberg \cite{nirenberg}), 
when $\lambda$ crosses $\lambda_c$, 
the equations bifurcate from 
the trivial solution to a steady state solution in $H'$. Therefore 
the attractor $\Sigma_\lambda$
contains at least one steady state solution. 
Secondly, it's easy to check that the equations (\ref{eq2.4})-(\ref{eq2.7}) are
translation invariant in x-direction. Hence if 
 $\psi_0(x,z)=(U(x,z),T(x,z),S(x,z))$ is a steady state solution, 
 then $\psi_0(x+\rho,z)$ are steady state solutions as well. By periodic 
 conditions in x-direction, the set 
\begin{align*}
S_{\psi_0}=\{\psi_0(x+\rho,z) | \rho \in \mathbb R \}
\end{align*}    
is a cycle  homeophic to $S^1$ in $H_{1}$. Therefore the steady state of 
 (\ref{eq2.4})-(\ref{eq2.7}) generates a cycle of steady state solutions.
Hence the bifurcated attractor $\Sigma_\lambda$ contains at least a cycle
of steady state solutions.
\subsection{Asymptotic structure of solutions}
It's easy to see that for any initial value 
$\psi_0 =(U_0, T_0, S_0) \in H$, 
there is a time $t_1 \ge 0$ such that 
the solution $\psi =(U(t, \psi_0), T(t, \psi_0), S(t, \psi_0))$   
is $C^\infty$ for $t> t_1$, and is uniformly bounded 
in $C^r$-norm for any given $r\ge 1$. 
Hence, by Theorem~\ref{th3.6}, we have 
\be
\label{eq6.1}
\lim_{t \to \infty} \min_{\phi \in \Sigma_\lambda} \| \psi(t, \psi_0) - \phi\|_{C^r} 
=0.
\ee
We infer then from (\ref{eq5.12})  and (\ref{eq5.14}) that 
for any steady state solution $\phi=(e, T, S) \in \Sigma_\lambda$ of 
(\ref{eq2.4})-(\ref{eq2.7}), the vector field $e=(e_1, e_2)$ can be expressed 
as 
\be
\label{eq6.2}
\left\{
\begin{aligned}
& e_1 =- r \sqrt{2}\sin (\alpha x+\theta) \cos \pi z + 
    v_1(x_{111}, y_{111},\beta_{111}), \\
& e_2 = r\cos (\alpha x+\theta) \sin \pi z +
    v_2(x_{111}, y_{111},\beta_{111}) , 
\end{aligned}\right.
\ee
for some $0\le \theta \le 2\pi$. Here 
\be
\label{eq6.3}
\left\{
\begin{aligned}
& r=\sqrt{x_{111}^2 +y_{111}^2 } = \sqrt{\frac{8\beta_{111}(\lambda)}{\delta}} 
      + o(\sqrt{\beta_{111}(\lambda)}) 
    &&\text{ if } \lambda > \lambda_c, \\
& v_i(x_{111}, y_{111}, \beta_{111}) = o(\sqrt{\beta_{111}(\lambda)}) 
     && \text{ for }i=1, 2.
\end{aligned}\right.
\ee

Now we show that the vector field 
\be
\label{eq6.4}
e_0 = \left(- r \sqrt{2}\sin \alpha x \cos \pi z , 
r\cos \alpha x \sin \pi z  
  \right)
\ee
is regular in $\Omega=\mathbb R^1 \times (0, 1)$. The singular 
points of $e$ are 
$(x,z)=((k+\frac{1}{2})\sqrt{2},\frac{1}{2})$, $(k\sqrt{2},1)$ and
$(k\sqrt{2},0)$ with $k\in \mathbb Z$, and  
\begin{align*}
 det De_0(x,z) & = det \left(\begin{matrix}
-\sqrt{2}r\alpha\cos \alpha x\cos\pi z & \sqrt{2}r\pi\sin\alpha x \sin \pi z\\
-r\alpha\sin \alpha x \sin \pi z & r\pi\cos \alpha x \cos \pi z 
\end{matrix}\right)\\
    &=\left\{\begin{aligned} 
& r^2\pi^2\ne 0, \,\, \text{for} \,\, 
    (x,z)=((k+\frac{1}{2})\sqrt{2},\frac{1}{2}),\\ 
& -r^2\pi^2\ne 0, \,\, \text{for} \,\,
    (x,z)=(k\sqrt{2}, 0),\, (k\sqrt{2}, 1).
\end{aligned}\right.
\end{align*}
Therefore the vector field (\ref{eq6.4}) is regular, and consequently,
the vector field $e$ in (\ref{eq6.2}) is regular for any 
$\lambda_c < \lambda < \lambda_c+\epsilon$ for some $\epsilon$ small. It 
follows from Theorem~\ref{th3.10} that the vector field $e$ of (\ref{eq6.2}) is
topologically equivalent to the vector field $e_0$ given by (\ref{eq6.4}), 
 which has the topological structure as shown in Figure~\ref{fg2.3}.  
\subsection{Proof of Theorem 3.5}
Inferring from (\ref{eq5.12})-(\ref{eq5.14}) and Lemma~\ref{le5.1},
the equations bifurcate from $(0,\lambda_c)$ to a repeller 
$\Sigma_{\lambda}^1$ for $\lambda<\lambda_c$, which is homeomorphic
to $S^1$. Since $(U,T,S)=(0,0,0)$ is a global attractor for 
$\lambda$ near $0$ , there 
exists a saddle-node bifurcation point $\lambda_0$ in between 
0 and $\lambda_c$. This completes the proof. 

\bibliography{2d-dd}

\end{document}